\documentclass[twocolumn]{aa}
\usepackage{graphicx}
\usepackage{txfonts}          
\usepackage[]{hyperref}
 \usepackage{graphicx}
 \usepackage[normalem]{ulem}
 \useunder{\uline}{\ul}{}
 \usepackage{lscape}
 \usepackage{xcolor}
 \usepackage{float} 
 \usepackage[colorinlistoftodos]{todonotes}
 \usepackage{amssymb}
 \usepackage{amsmath}
  \usepackage{booktabs}

\begin{document}

    \title{Multi-spacecraft constraints on relativistic solar energetic particle transport in the widespread 28 October 2021 event}

    \author{
        E. Lavasa \inst{1,2,3}
        \and
        J.~T. Lang\inst{4}
        \and
        A. Papaioannou \inst{3}
        \and
        R.~D. Strauss \inst{5,6}
        \and
        S. ~A. Mallios \inst{3}
        \and 
        A. Hillaris \inst{1}
        \and
        A. Kouloumvakos \inst{7}
        \and
        A. Anastasiadis \inst{3}
        \and
        I. A. Daglis \inst{1}
        }    
    
    \institute{
        National \& Kapodistrian University of Athens, Athens, Greece
            \email{elavasa@phys.uoa.gr}
        \and
        Institute for the Management of Information Systems, "Athena" Research Center, Marousi, Greece
       \and
        Institute for Astronomy, Astrophysics, Space Applications and Remote Sensing (IAASARS), National Observatory of Athens, I. Metaxa \& Vas. Pavlou St., 15236 Penteli, Greece
         \and
    Space Research Laboratory, University of Turku, Finland
     \and
    Centre for Space Research, North-West University, Potchefstroom, South Africa
    \and
    National Institute for Theoretical and Computational Sciences (NITheCS), Potchefstroom, South Africa
     \and
    The Johns Hopkins University Applied Physics Laboratory, 11101 Johns Hopkins Road, Laurel, MD 20723, USA
            }

\titlerunning{The 28 October 2021 multi-spacecraft SEP event}  
%   \date{Received September 15, 1996; accepted March 16, 1997}

% \abstract{}{}{}{}{} 
% 5 {} token are mandatory
 
\abstract
  % context heading (optional)%                 aa.dem
   {}
   {We investigated the transport of solar energetic particles (SEPs) during the relativistic widespread event of 28 October 2021, quantifying the role of parallel and perpendicular diffusion and constraining the spatial extent of the injection region.}
   {We employed inverse modeling of particle focused transport and 2D numerical simulations including cross-field diffusion. Multi-spacecraft observations from STEREO-A, Solar Orbiter, and near-Earth spacecraft are used to reproduce particle intensity profiles and anisotropies across a wide range of electron and proton energies. Simulated flux profiles are compared across different heliolongitudes to derive consistent transport parameters.}
   {The analysis yields parallel mean free paths within or slightly above the Palmer consensus range, and perpendicular mean free paths that correspond to $\sim$1–3\% of parallel for electrons and $\sim$5–10\% for protons. The injection region is found to be relatively narrow ($\leq$20$^{\circ}$), and decreasing with particle rigidity. Multipoint simulations indicate that the observed flux and anisotropy profiles can only be reproduced by a narrow injection region and significant cross-field diffusion. Electron and proton release times align well with the parent X1.0 flare and associated coronal mass injection (CME) onset, indicating that a compact acceleration region coupled with efficient interplanetary diffusion governed the event’s broad spatial extent.}
   {}


   \maketitle
%
%________________________________________________________________

\section{Introduction} \label{sec:intro}
 %%%%%%%%%%%%%%%%%%%%%%%%%%%%%%%%%%%%%%%%%%%%%%%%%%%%%%%%%%%%%%%%%%%%%%%%%%%
\begin{figure*}[h!]
\centering
\includegraphics[width=0.65\textwidth]{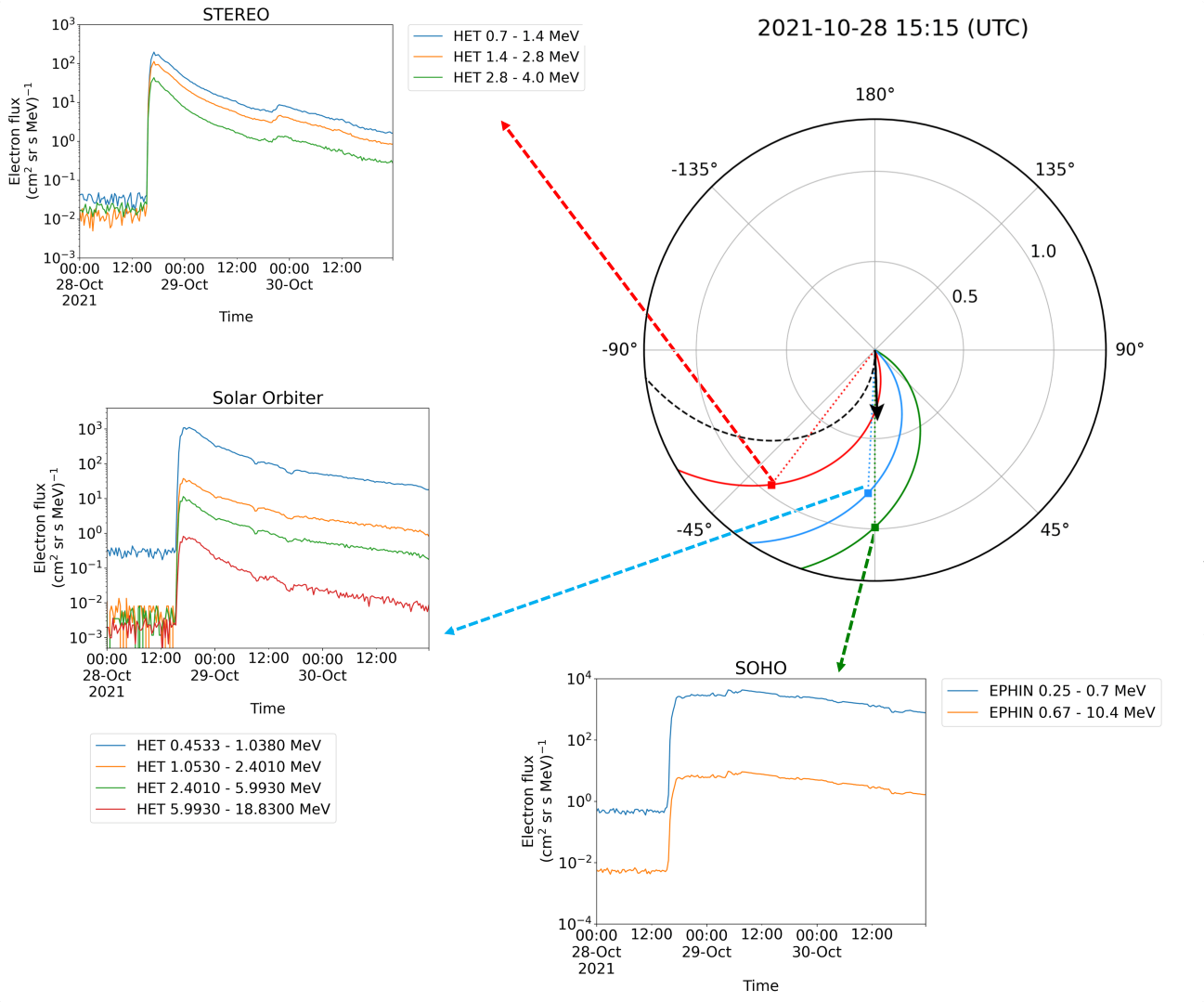}
\caption{View of the ecliptic plane (top right) from solar north showing the positions of various spacecraft, and their respective Parker spirals, on 28 October 2021 at 15:15~UT. The black arrow and dashed black curve show the flare reference location and corresponding Parker spiral. The insets show the electron flux intensity time series for each corresponding spacecraft. The top right figure is derived from the Solar MAgnetic Connection Haus tool \citep[\url{https://solar-mach.github.io/}; ][]{Gieseler_2023}.} 
\label{fig:sc_pos}
\end{figure*}
%%%%%%%%%%%%%%%%%%%%%%%%%%%%%%%%%%%%%%%%%%%%%%%%%%%%%%%%%%%%%%%%%%%%%%%%%%%%%%
Solar energetic particle (SEP) events are sudden increases in particle radiation from the Sun following eruptive flares and coronal mass ejections (CMEs) \citep{2016LRSP...13....3D, anastasiadis2019,2021LNP...978.....R}. When solar protons are accelerated to relativistic energies ($>$500~MeV), secondary neutrons produced in atmospheric cascades can be detected by ground-based neutron monitors, marking a ground level enhancement (GLE) \citep[e.g.,][]{macau_mods_00003837}. SEPs—mainly protons and electrons, but also heavier ions—pose significant hazards to astronauts, aviation, and space infrastructure, motivating ongoing research into their acceleration and transport \citep{Whitman_2023}.

Solar energetic particle acceleration from $\sim$10~keV to $\sim$GeV energies is primarily attributed to two mechanisms: direct current (DC) acceleration during magnetic reconnection in flares and diffusive shock acceleration at CME-driven shocks \citep{Vainio2018}. Although both processes are supported by observations, their relative contributions remain uncertain \citep{klein2017,anastasiadis2019}. After acceleration, SEP propagation through interplanetary space is governed by their motion along the heliospheric magnetic field (HMF) lines, often modeled as a Parker spiral\citep{Parker1958}. Magnetic turbulence causes pitch-angle scattering, quantified by the parallel mean free path ($\lambda_{\parallel}$), while cross-field transport is quantified by the perpendicular mean free path ($\lambda_{\perp}$) \citep{Strauss_2017,Laitinen_2019}. Modern models such as PARADISE \citep{Wijsen_2020} couple CME-driven shock acceleration with 3D transport to reproduce multi-spacecraft SEP observations. Recent studies confirm a rigidity-dependent valley in $\lambda_{\parallel}$ \citep{Lang_2024,Droge_2000}, with values generally within the Palmer consensus range (0.08–0.3 AU at 1~AU) but varying significantly between events \citep{Palmer_1982,Droge_2009,Engelbrecht_2022}.

Widespread SEP events, sometimes spanning $>$180° in longitude \citep{Gomez-HerreroEA_2015, DresingEA_2025}, indicate additional transport processes beyond simple field-aligned motion. Proposed mechanisms include cross-field diffusion and magnetic turbulence \citep{Strauss_2017,Steyn_2020}, field-line meandering \citep{laitinen2016}, extended injection sources \citep{dalla_2020,Koulou2024}, and interaction with the heliospheric current sheet (HCS) \citep{Battarbee_2018,2025ApJ...991..104W}. Models incorporating perpendicular diffusion reproduce observed spreads with $\lambda_{\perp}/\lambda_{\parallel}$ ratios of a few percent at 1~AU \citep{Dresing_2014,Droge_2014}. Open questions remain regarding the true magnitude and variability of $\lambda_{\perp}$, and the relative roles of coronal and interplanetary processes in shaping SEP distributions.

For this work we investigated the relativistic SEP event of  28 October 2021 (GLE73) to constrain parallel and perpendicular diffusion for both protons and electrons. Using numerical simulations, we estimated the rigidity dependence of $\lambda_{\parallel}$ and $\lambda_{\perp}$ and the extent of the injection region. Section~\ref{sec:2} presents the observational overview, Section~\ref{sec:3} describes the modeling approach, and Sections~\ref{sec:4} and \ref{sec:5} discuss the results and their implications.

%%%%%%%%%%%%%%%%%%%%%%%%%%%%%%%%%%%%%%%%%%%%%%%%

\section{Event overview, instruments, and data}
\label{sec:2}
\subsection{Overview of the SEP event} 

The intense SEP event of 28 October 2021, produced a marked enhancement in neutron fluxes observed at multiple ground-based neutron monitors, and was classified as the GLE73 event \citep[][and references therein]{Papaioannou2022, Mishev2023}. A strong X1.0 class flare and fast halo CME have been identified as the drivers of particle acceleration, accompanied by type II, III, and IV radio signatures implying shock propagation, energetic particles escaping into interplanetary (IP) space along open magnetic field lines, and trapped energetic particles in closed coronal structures, respectively \citep{Klein_2022, Papaioannou2022}. Particles were accelerated to relativistic energies, with the highest rigidities recorded at $\sim$3~MV for electrons at STEREO-Ahead and $\sim$2.4~GV for protons at Earth using neutron monitor data. This is a perfect candidate event to study SEP focused transport and perpendicular diffusion, with particle flux enhancements detected by multiple spacecraft, including STEREO-Ahead \citep[STA;][]{2008SSRv..136....5K}, Parker Solar Probe \citep[PSP;][]{Fox2016}, Solar Orbiter \citep[SolO;][]{Muller2020}, the Solar and Heliospheric Observatory \citep[SOHO;][]{1995SoPh..162....1D}, and Geostationary Operational Environmental Satellites \citep[GOES;][]{grubb1975sms} at different longitudinal separations from the eruption site on the Sun.  

The highest fluxes were recorded at STA which had the best magnetic connection to the source region, while E$>$150~MeV protons were recorded on Mars by the Radiation Assessment Detector (RAD) of the Mars Science Laboratory (MSL) experiment, mounted on the Curiosity rover \citep[see Fig.~1 of][]{Koulou2024}. The detection of high-energy particles at Mars, which was situated almost diametrically opposite the flare site, and therefore magnetically disconnected from it, points to highly effective perpendicular diffusion during this event. Comprehensive analyses of proton and solar context observations are presented in the works of \citet{Papaioannou2022}, \cite{Klein_2022}, \cite{Mishev2023},  and \citet{Koulou2024}. An overview of observed electron fluxes at the different locations is presented in Fig.~\ref{fig:sc_pos}. PSP was excluded from the analysis due to data gaps that made using these data challenging (see also Fig.~1 in \citet{Koulou2024}). 

%Observations show undisturbed magnetic fields in all three locations, without the presence of large magnetic structures. Rather stable plasma and solar wind conditions are observed at STA and the Earth, with small fluctuations. The corresponding information on plasma and solar wind is not available at Solar Orbiter at the time of the event, due to data gap in the Solar Wind Analyzer (SWA) instrument. A small shock-like structure is observed at Earth but is negligible and not included in IP shock lists. In the appendix we provide particle data along with context observations on the magnetic field and plasma properties (where available) at STA, SolO and the Earth, to complement the event overview from an observational perspective.

\subsection{Data and instruments}

In the analysis we used data from five missions: STA, SolO, Wind, SOHO, and GOES. At the time of the event, the STA spacecraft was located at a heliocentric distance of approximately 0.96~AU with a magnetic connection angle (CA) of about 25$^{\circ}$ relative to the parent solar activity region. Particle flux measurements from two instruments  on board STA were analyzed. From the High Energy Telescope (HET), proton intensities were obtained over 11 energy channels spanning mean energies of 14.3--77.5~MeV. Electron measurements were provided by both the Solar Electron Proton Telescope (SEPT) and HET: SEPT electrons were used in the energy range of 0.05--0.4~MeV (15 differential channels), while HET contributed higher-energy electron observations between 1--3.3~MeV (3 channels). In addition to fluxes, pitch-angle distribution (PAD) data of electrons from STA/SEPT were also processed to examine directional anisotropies.

SolO was situated closer to the Sun than STA and the Earth, at a heliocentric distance of 0.80~AU with a CA of roughly 45$^{\circ}$. Proton observations were obtained from the penetrating particle channels of SolO/HET, covering discrete mean energies of 108, 130, 300, and 896~MeV \citep[see also][and references therein]{Koulou2024}. Electron intensities were drawn from two complementary sources: the Electron Proton Telescope (EPT), which provided measurements at 0.115–0.4~MeV across four channels, and SolO/HET, which supplied electron data at 1.0 and 2.5~MeV. 

From the near-Earth vantage point, located at 0.99~AU with a CA of approximately 81$^{\circ}$, a combination of particle instruments was employed. The Wind / 3D Plasma Analyzer (3DP) instrument contributed electron observations in the range of 0.110–0.310~MeV (3 channels). Complementary higher-energy electron data were provided by SOHO/EPHIN, spanning 0.42–2.64~MeV (2 channels). For protons, we made use of measurements from the GOES / Space Environment In-Situ Suite (SEISS), specifically at 108 and 130~MeV.

In several instances, to facilitate inter-spacecraft comparison, the effective mean particle energies were approximated by combining or binning adjacent nominal instrument channels. This approach ensured consistent energy coverage across datasets, while minimizing the effects of individual channel noise or resolution limitations. 

Magnetic field and plasma measurements were obtained for the three observers, where possible, through the SOLER Multi-instrument plot tool\footnote{\url{https://github.com/soler-he/sep_tools}} \citep{gieseler_2025_15827561}. The In-situ Measurements of Particles and CME Transients (IMPACT) / Magnetometer (MAG) and the PLAsma and SupraThermal Ion Composition (PLASTIC) instruments on board STA are used by the tool to obtain magnetic field and plasma measurements, respectively. At SolO, magnetic field data were extracted by the MAG instrument, while plasma measurements were unavailable due to a data gap in the Solar Wind Analyzer (SWA). At Earth, the tool extracted magnetic field measurements from Wind / Magnetic Field Investigation (MFI) and plasma properties by Wind/3DP.

\section{Methods}
\label{sec:3}
\subsection{Focused transport}

In the absence of large-scale heliospheric disturbances, such as CMEs or interplanetary shocks, the HMF is often nicely approximated by a smooth large-scale structure. This large-scale field can be described by the Parker spiral (an Archimedean spiral configuration), which results from the outward flow of the solar wind combined with the rotation of the Sun \citep{Parker1958}. Superposed on this average structure is a fluctuating component associated with magnetic turbulence.

Charged particles injected into the heliosphere propagate through this environment under the combined influence of the large-scale HMF and its turbulent variations. Their motion can be understood as consisting of two principal components. First, particles undergo adiabatic guiding-center motion along the average Parker spiral field, which governs their large-scale transport from the Sun into interplanetary space. Second, the presence of magnetic turbulence leads to pitch-angle scattering, a process in which the particle’s direction of motion relative to the mean magnetic field is randomized through interactions with turbulent magnetic fluctuations. The interplay between these two processes determines the efficiency and characteristics of particle propagation, influencing both their longitudinal spread and their temporal profiles at different heliocentric distances. The evolution of the phase space density of a particle population in the heliosphere is typically treated quantitatively within the framework of focused transport theory. The governing equation is a form of the focused transport equation \citep[FTE; ][]{roelof1969}, which can be viewed as a generalization of the Fokker–Planck equation adapted to a non-uniform, diverging HMF. Equation~\ref{eq:1} incorporates key physical processes (streaming and focusing along the large-scale Parker spiral field, pitch-angle diffusion due to magnetic turbulence) in a simplified expression where advection and energy losses are not considered:

\begin{equation}
    \frac{\partial f}{\partial t} + \mu v \frac{\partial f}{\partial s} + \frac{1}{2L(s)}\frac{\partial}{\partial \mu}\left[(1 - \mu^2)v f \right] = \frac{\partial}{\partial \mu} \left[ D_{\mu\mu}\frac{\partial f}{\partial \mu} \right] .
    \label{eq:1}
\end{equation}

Here \textit{f(s, $\mu$, t)} is the gyrotropic distribution function of particles at a certain energy in a given flux tube; \textit{v} their constant speed, which is assumed to be much higher than the solar wind speed; \textit{s} the field-aligned coordinate; \textit{L(s)} the focusing length of the magnetic field; and \textit{$D_{\mu \mu}$} the pitch-angle diffusion coefficient. The terms, from left to right, correspond to temporal change in the distribution function, spatial change due to particle streaming along the magnetic field (advection term), pitch-angle changes due to adiabatic focusing (focusing term), and pitch-angle scattering. 
This reduced expression is derived under the assumption that spatial variations in the large-scale HMF occur on scales much larger than the particles' gyro-radius so that  the focusing length is much larger than the gyro-radius. This condition breaks very close to the Sun (low corona) or in the presence of large magnetic field gradients as in CMEs or current sheets. 

In this treatment, it is assumed that spatial variations perpendicular to the mean magnetic field are negligible so that cross-field particle transport is removed. In other words, the particle distribution is considered to be uniform across neighboring magnetic flux tubes, such that the solution for the phase space density is effectively one-dimensional along the guiding field line. This simplifying assumption implies that adjacent flux tubes are populated identically, and that the evolution of the distribution function depends solely on the distance along the interplanetary magnetic field, pitch angle, and time. While this approximation neglects perpendicular diffusion or drifts of the particles away from their original field lines, it provides a tractable framework for investigating the essential physics of particle streaming and scattering along the Parker spiral.

The 1D SEP-propagator model\footnote{\url{https://github.com/RDStrauss/SEP\_propagator}} \citep{vandenberg2020} was used to simulate particle fluxes at the magnetically well-connected observer, i.e., STA. This is a numerical solution, based on a finite difference scheme, that solves the FTE under the assumption of specific functional forms of the injection function (delta-like or time-dependent), HMF (Parker spiral with a turbulent component), and pitch-angle diffusion coefficient $D_{\mu \mu}$ driven from small-scale magnetic turbulence in the HMF. The diffusion coefficient is parameterized by the particles' parallel mean free path, $\lambda_{\parallel}=3\kappa_{\parallel}/v$ in relation to the parallel diffusion coefficient $\kappa_{\parallel}$ \citep{bieber_1994, Shalchi2009}: 

\begin{equation}
    \lambda_{\parallel} = \frac{3}{8}v \int_{-1}^{1} \frac{(1 - \mu^2 )^2}{D_{\mu \mu}(\mu)}d\mu.
    \label{eq: lambda_paral}
\end{equation}

Assuming a magnetic slab turbulence spectrum, so that pitch-angle scattering is dominated by resonant interactions between particles and magnetic fluctuations propagating parallel to the mean HMF \citep{Matthaeus1990}, the functional form of $D_{\mu \mu}$ related to the slab turbulence spectrum under quasi-linear theory (QLT) approximation \citep{Shalchi2009} can be parameterized as \citep{Droge_2000, Lang_2024} 

\begin{equation}
    D_{\mu \mu} = D_0 \left( |\mu|^{q-1} + H \right) (1 - \mu )^2,
    \label{eq: Dmumu}
\end{equation}

where $q = 5/3$ is the Kolmogorov inertial range spectral index \citep{Kolmogorov1941}, $H=0.05$ is an arbitrary constant to represent dynamical effects with value selected to give finite scattering at $\mu = 0$, and $D_0$ is the scattering amplitude defined by normalization to $\lambda_{\parallel}$ \citep{vandenberg2020, Lang_2024}. We note that more advanced treatments, such as the second-order quasi-linear approximation discussed by \cite{Shalchi2009}, account for higher-order effects and can  also be used in strong turbulence with steep spectra.  
A time-dependent particle injection following a Reid--Axford profile is assumed at the inner boundary $s_0$ which is set to 0.05 AU: 

\begin{equation}
    f_0 (s_0 , t) = \frac{C}{t} \exp{\left[ -\frac{t_a}{t} - \frac{t}{t_e} \right]}.
    \label{eq: injection}
\end{equation}

The acceleration and escape times of the particles are defined as $t_a$ and $t_e$, respectively, in Equation~\ref{eq: injection}, and $C$ is a constant.  
We inverse-modeled the event at various particle energies (rigidities) by fitting simulated to observed averaged intensity profiles, until satisfactory agreement was achieved. Based on the best-fit solution, we inferred $\lambda_{\parallel}$ as well as the acceleration and escape time ($t_a, t_e$) of the particles at each species and energy. The process included (i) the temporal alignment of onsets, with the observed onsets defined using the $3\sigma$ onset determination method \citep{palmroos_2022}, and  (ii) the scaling of simulated intensities to the observed (mean) peak intensities and background levels \citep{Lang_2024}.  The temporal resolution of the simulations was $\sim$1~minute, so particle fluxes were time-averaged at a 1 minute cadence as well. The exception was in the treatment of $> 100$ MeV protons in SolO penetrating channels, where we applied 5 minute temporal averaging due to data cadence. The goodness of fit was measured by the coefficient of determination \citep[$R^2$; ][]{Achen_1982,Berry_Feldman_1985}. 

Simulated anisotropies for STA/SEPT electrons were additionally compared to the observed ones, by visual inspection, to further validate the optimal fits. The observed first-order anisotropies were calculated with the weighted sum method,

\begin{equation}
    A = 3\frac{\int_{-1}^{1} \mu \, f \, d\mu }{\int_{-1}^{1}  f \, d\mu} \approx 
    3 \frac{\sum_{1}^{n} \; \mu_{i} \; F_{i}}{\sum_{1}^{n} \; F_{i} } = 
    3 \frac{\sum_{1}^{n} \; \mu_{i} \; I(\mu_{i}) \; \delta \mu_{i}}{\sum_{1}^{n} \; I(\mu_{i}) \; \delta \mu_{i}} ,
    \label{eq:2}
\end{equation}

where the sums are over the number \textit{n} of sectors, $\mu_{i}$ is the mean pitch-angle cosine, $\delta \mu_{i}$ is the pitch-cosine range, and I($\mu_{i}$) is particle intensity at a given sector. Pitch angles and particle fluxes from the four STA/SEPT sectors (sun, antisun, north, south) were used, and the result was smoothed with size-2 window moving average. The evolution of anisotropy for 115 and 314~keV electrons during the event was further investigated, with the comparison of simulated PADs at different time steps to the actual ones. In particular, simulated PADs at 0.5, 0.75, 1, 1.25, 1.5, 2, 3, and 5~hours after the injection were compared to the normalized peak intensities detected at the same time at the four different STA/SEPT sectors and corresponding mean pitch angles. 

The 1D model was further used to estimate $\lambda_{\parallel}$, $t_{\alpha}$, and $t_{e}$ for particles detected at the magnetically disconnected observers (i.e., SolO, Earth) at higher proton energies than the range of STA/HET (maximum energy $\sim$77.5~MeV). These estimates served as initial constraints on the range of focused transport parameters employed in the subsequent two-dimensional (2D) simulations, that included cross-field diffusion. By narrowing the parameter space to physically reasonable values, the computational effort required for the simulations was significantly reduced, thereby rendering the search for optimal 2D transport parameters both tractable and efficient.

\subsection{Cross-field diffusion} \label{sec:rs}

Particle transport under adiabatic focusing and pitch-angle scattering, also considering perpendicular diffusion to the HMF and neglecting energy losses, is described by the following transport equation (TE) \citep{Skilling_1971}:

\begin{multline}
   \frac{\partial f(\boldsymbol{x}, \mu, t)}{\partial t} = -\boldsymbol{\nabla} \cdot (\mu v \boldsymbol{\hat{b}}f) - \frac{\partial}{\partial \mu} \left( \frac{1-\mu^2}{2L} v f \right) \\ + \frac{\partial}{\partial \mu} \left( D_{\mu \mu} (\boldsymbol{x}, \mu) \frac{\partial f}{\partial \mu} \right) + \boldsymbol{\nabla} \cdot \left( \boldsymbol{D}_{\perp}^{(x)} (\boldsymbol{x}, \mu ) \cdot \boldsymbol{\nabla}f \right).
   \label{eq:3}
\end{multline}

The terms from left to right correspond to the temporal evolution of the particle's distribution function, particle streaming along the HMF indicated by the unit vector $\boldsymbol{\hat{b}}$, adiabatic focusing, pitch-angle scattering, and cross-field diffusion. The 2D SEP-propagator model\footnote{\url{https://github.com/RDStrauss/2D-SEP-model}} \citep{Strauss_2015} numerically solves the TE in the ecliptic plane (radial / heliocentric distance \textit{r} and heliographic longitude $\phi$), based on a finite difference scheme. The effect of magnetic turbulence is modeled in the diffusion coefficients, assuming the HMF is composed of a locally uniform Parker field and a turbulent component with slab and 2D counterparts \citep[see, e.g.,][]{1990JGR....9520673M} the spectra of which can be configured in the code. The slab component is related to the pitch-angle diffusion coefficient $D_{\mu \mu}$ parameterized by $\lambda_{\parallel}$ as in the 1D model (Equations~\ref{eq: lambda_paral}-\ref{eq: Dmumu}). Cross-field diffusion is controlled by the perpendicular diffusion coefficient $D_{\perp}$ for which we select the field line random walk (FLRW) approximation where cross-field particle transport is governed primarily by the stochastic wandering of magnetic field lines \citep{Qin2014, Strauss_2015}:

% \begin{equation}
%     D_{\perp}(\mu) = \frac{4}{\pi} D_{\perp, 0} \sqrt{1 - \mu^2}
%     \label{eq: Dperp}
% \end{equation}

\begin{equation}
    D_{\perp}(\mu) = 2 D_{\perp, 0} |\mu|.
    \label{eq: Dperp}
\end{equation}

The value of $D_{\perp,0}$ is calculated from the specified perpendicular mean free path $\lambda_{\perp}$,

\begin{equation}
    \lambda_{\perp} = 3 \frac{\kappa_{\perp}}{v},
    \label{eq: lambda_perp}
\end{equation}

where $\kappa_{\perp}$ is the isotropic perpendicular diffusion coefficient: 

\begin{equation}
    \kappa_{\perp} = \frac{1}{2} \int_{-1}^{1}D_{\perp}(\mu) d\mu .
    \label{eq:kappa}
\end{equation}

The heliographic longitude $\phi_0$ and angular size of the injection source $\Delta\Phi$, as well as observers at different radial distances and heliographic longitudes, can be configured to simulate a given event. We assume a Gaussian in $\phi$, time-dependent injection at the inner boundary $r_0 = 0.05$ AU at $\phi = \phi_0$ with angular size $\Delta\Phi$ expressed as 

\begin{equation}
    f(r, \phi, t) = \frac{C_2}{t} \exp{\left[ -\frac{t_a}{t} - \frac{t}{t_e} \right]} \cdot \exp{-\frac{(\phi - \phi_0)^2}{\Delta\Phi ^2}}  .
\end{equation}

Adiabatic energy losses and co-rotation effects are neglected; even so, extended cross-validation analysis shows that the results of this 2D finite difference-based SEP transport model are in very good agreement with the results from full 3D simulations using stochastic differential equation (SDE) schemes \citep{Steyn_2020}.   

After configuring the locations of our observers (STA, SolO, and Earth) as well as the helio-longitude of the injection source using the longitude of the parent X1.0 flare (W02), we fitted simulated-to-observed profiles at all available observers to inverse-solve for the particles’ $\lambda_{\perp}$ and the size of the injection region. We used $\lambda_{\parallel}$, acceleration, and escape times constrained by the 1D inverse-solutions. 

Recordings at all three observers at similar rigidities were available only for the electrons. We avoided channel binning when nominal channels with similar energies existed; for example, we simulated the transport of electrons at 115~keV to fit STA/SEPT 115~keV [105--125~keV], SolO/EPT 116~keV [111--121~keV], and Wind/3DP 110~keV [76--141~keV]. When large differences were encountered between nominal channels of the different observers, we averaged neighboring channels together if the resulting energy range was not too large. For example, we simulated 1~MeV electrons to compare with STA/HET 0.99~MeV [0.7--1.4~MeV] and SolO/HET 1.04~MeV [0.45--2.4~MeV] (channels 0--1). For the high-energy protons >100~MeV, SolO/HET and GOES/SEISS at the Earth measure similar energies (108 and 130~MeV), while for the penetrating channels no common ground could be found in the available instruments so we chose to model SolO/HET 300 and 896~MeV to extend the analysis to the highest rigidity recorded during the event by space-borne instruments. 

Simulated-to-observed anisotropies were further compared at the three observers, STA, SolO, and Earth, at three electron energies detected with multiple sectors by STA/SEPT (4 sectors), SolO/EPT (4), and Wind/3DP (8), namely at 115, 180, and 314~keV. Anisotropies were also calculated for the 108 and 130~MeV proton channels of SolO/HET using four-sector data.              

%%%%%Begin table%%%%%%%
\begin{table}[htbp]
\footnotesize
     % \centering
     \caption{Particle energies and instruments used in the 2D transport analysis.}
     \setlength{\tabcolsep}{0.9\tabcolsep}
    \begin{tabular}{ c c c c c}
    \toprule
    \textbf{Simulation} & \textbf{Observer /} & \textbf{Mean} & \textbf{Energy} & \textbf{Onset} \\
    \textbf{energy} & \textbf{Instrument} & \textbf{energy} & \textbf{range} &  \\
    \textbf{(MeV)} &  & \textbf{(MeV)} & \textbf{(MeV)} & \textbf{UT} \\
    \hline
    \multicolumn{5}{c}{\textbf{Electrons}} \\
    \hline
    0.115 & STA/SEPT & 0.115 & 0.105-0.125 & 15:41 \\
    & SolO/EPT & 0.116 & 0.111-0.121 & 15:52 \\
    & Wind/3DP & 0.110 & 0.076-0.141 & 16:10\\
    \hline
    0.180 & STA/SEPT & 0.179 & 0.165-0.195 & 15:41\\
    & SolO/EPT & 0.177 & 0.169-0.185 & 15:51 \\
    & Wind/3DP & 0.182 & 0.127-0.236 & 16:09 \\
    \hline
    0.314 & STA/SEPT & 0.314 & 0.295-0.335 & 15:43 \\
    & SolO/EPT & 0.321 & 0.307-0.336 & 15:46 \\
    & Wind/3DP & 0.310 & 0.217-0.402 & 16:08 \\
    \hline
    0.400 & STA/SEPT & 0.399 & 0.375-0.425 & 15:42 \\
    & SolO/EPT & 0.417 & 0.399-0.435 & 15:51 \\
    & SOHO/EPHIN & 0.420 & 0.250-0.700 & 15:54 \\
    \hline
    1.0 & STA/HET & 0.99 & 0.7-1.4 & 15:40 \\
    & SolO/HET & 1.04 & 0.45-2.4 & 15:41 \\
    \hline
    2.5 & STA/HET & 2.4 & 1.4-4.0 & 15:39 \\
    & SolO/HET & 2.5 & 1.05-5.99 & 15:40 \\
    & SOHO/EPHIN & 2.6 & 0.67-10.4 & 15:55 \\
    \hline \hline
    \multicolumn{5}{c}{\textbf{Protons}} \\
    \hline
    108 & SolO/HET & 108 & & 16:05 \\
    & GOES/SEISS & 108 & 99-118 & 16:20 \\
    \hline
    130 & SolO/HET & 129.6 & & 15:50 \\
    & GOES/SEISS & 133 & 118-150 & 16:25 \\
    \hline
    300 & SolO/HET & 300 & & 15:45 \\
    \hline
    896 & SolO/HET & 896 & & 15:50 \\
    
    \bottomrule

    \end{tabular}
\label{tab:tab1}
\end{table}
%%%%end table%%%
\subsection{Transport conditions}
We examined the magnetic field and plasma observations at the three vantage points used in the transport analysis—STA, SolO, and Earth—as shown in Fig.~\ref{fig: field_and_plasma} and detailed in Appendix~\ref{appenA}. Specifically, Fig.~\ref{fig: field_and_plasma} presents particle measurements (top panels) together with the corresponding magnetic field (middle) and plasma data (bottom panels), where available.

Across all vantage points, the magnetic field observations indicate generally undisturbed conditions with no evidence of large-scale magnetic structures. Plasma and solar wind parameters at STA and near Earth remain stable, exhibiting only minor fluctuations. For SolO, plasma and solar wind measurements are unavailable during the event due to a data gap in the SWA instrument. At Earth, a weak enhancement is present around the time of the electron event onset, although it is minor and is not included in the IP shock catalogs.\footnote{\url{https://ipshocks.helsinki.fi/}}

\section{Results}
\label{sec:4}
\subsection{Magnetically connected observer} \label{sec:magn_con}
\subsubsection{Rigidity dependence of transport parameters}

\begin{figure}[h!]
\centering
\includegraphics[width=0.48\textwidth]{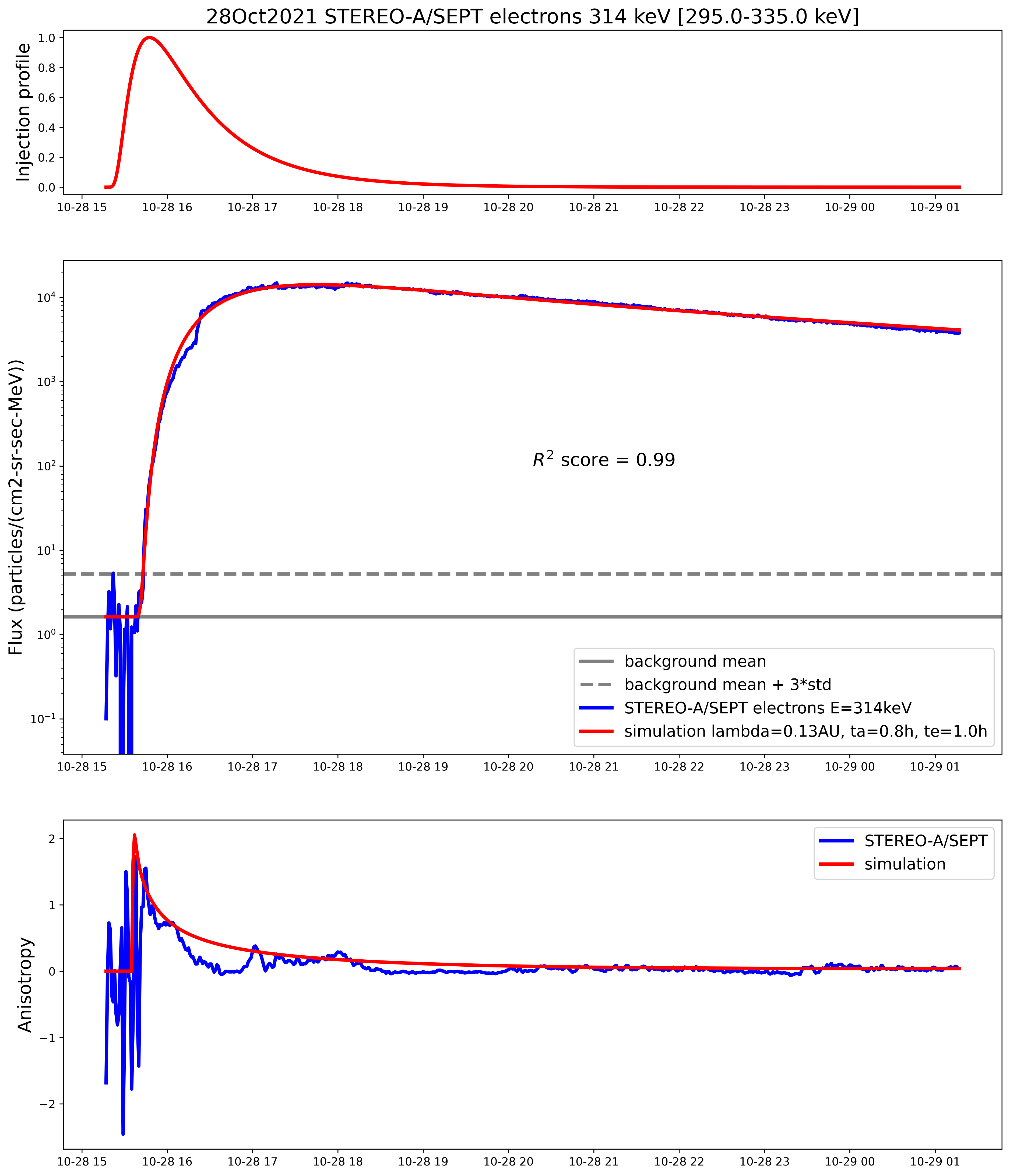}
\caption{Inverse solution for STA/SEPT electrons at 314~keV using the 1D SEP propagator model, for the particle flux (middle panel) and anisotropy (bottom panel) time profiles. The corresponding Reid-Axford injection profile is displayed in the top panel. The simulated and observed profiles are shown as red and blue lines, respectively. In the flux plot, the thick gray line corresponds to the average background and the dashed line to the average background  $+3\sigma$, the latter defining the cutoff for onset detection. The solution parameters ($\lambda_{\parallel}$, $t_a$, $t_e$, and  $R^2$ score) are also noted in the flux plot (middle panel). }  
\label{fig:e_314_1d}
\end{figure}

Inverse-solutions to constrain $\lambda_{\parallel}$, $t_a$, and $t_e$ were found for STA/HET protons [14.3--77.5~MeV, 11 channels], as well as STA/SEPT and HET electrons at energies [0.05--0.4~MeV, 15 channels] and [1--3.3~MeV, 3 channels], respectively. Results for both protons and electrons are listed in Table~\ref{tab:tab2}, including the optimal parameters and $R^2$ values. An example best-fit solution is displayed in Fig.~\ref{fig:e_314_1d} for STA/SEPT electrons at 314~keV, for the particle flux (top panel) and the anisotropy (bottom panel) time profiles. The $R^2$ scores are very high for electrons ($R^2$ = 96--99\%) and confirm the precision of derived parameter values. Scores are also high for protons 27.9--77.5~MeV ($R^2$ = 92--95\%), while moderate results ($R^2$ = 77--89\%) are retrieved for 14.3--25.1~MeV protons. 

%%%%%Begin table%%%%%%%
\begin{table}[!ht]
\footnotesize
     % \centering
     \caption{Inverse solutions for the magnetically connected observer, STEREO A.}
    \begin{tabular}{ c c c c c c}
    \toprule
    \textbf{Energy} & \textbf{Rigidity} & \textbf{$\lambda_{\parallel}$} & \textbf{$t_a$} & \textbf{$t_e$} & \textbf{$R^2$} \\
    \textbf{(MeV)} & \textbf{(MV)} & \textbf{(AU)} & \textbf{(hours)} & \textbf{(hours)} & \% \\
    \hline
    \multicolumn{6}{c}{\textbf{Electrons}} \\
    \hline
    0.05 & 0.232 & 0.08 & 0.5 & 0.6 & 97 \\
    0.06 & 0.255 & 0.07 & 0.6 & 0.4 & 97 \\
    0.07 & 0.276 & 0.12 & 0.6 & 2.0 & 97 \\
    0.08 & 0.297 & 0.12 & 0.6 & 2.0 & 98 \\
    0.94 & 0.324 & 0.13 & 0.6 & 2.0 & 98 \\
    0.115 & 0.361 & 0.14 & 0.8 & 1.8 & 98 \\
    0.135 & 0.395 & 0.15 & 0.8 & 1.8 & 98 \\
    0.155 & 0.427 & 0.17 & 0.9 & 2.0 & 98 \\
    0.179 & 0.463 & 0.14 & 0.8 & 1.4 & 99 \\
    0.210 & 0.508 & 0.14 & 0.8 & 1.4 & 99 \\
    0.240 & 0.550 & 0.16 & 0.8 & 1.6 & 99 \\
    0.274 & 0.595 & 0.14 & 0.8 & 1.2 & 99 \\
    0.314 & 0.647 & 0.13 & 0.8 & 1.0 & 99 \\
    0.354 & 0.697 & 0.14 & 1.0 & 1.0 & 99 \\
    0.399 & 0.752 & 0.13 & 0.8 & 1.0 & 99 \\
    1.0 & 1.42 & 0.12 & 1.6 & 0.4 & 97 \\
    2.0 & 2.46 & 0.12 & 1.6 & 0.4 & 97 \\
    3.3 & 3.78 & 0.15 & 2.0 & 0.4 & 96 \\
    \hline \hline
    \multicolumn{6}{c}{\textbf{Protons}} \\
    \hline
    14.3 & 164 & 0.26 & 12.2 & 0.9 & 77 \\
    16.0 & 174 & 0.29 & 12.2 & 0.9 & 83 \\
    18.1 & 185 & 0.30 & 12.2 & 0.9 & 80 \\
    22.2 & 205 & 0.35 & 12.2 & 1.0 & 83 \\
    25.1 & 218 & 0.38 & 11.5 & 1.0 & 89 \\
    27.9 & 231 & 0.34 & 9.4 & 1.0 & 92 \\
    31.4 & 245 & 0.33 & 5.0 & 1.4 & 94 \\
    34.6 & 257 & 0.34 & 6.0 & 1.2 & 94 \\
    38.0 & 269 & 0.34 & 8.0 & 0.9 & 94 \\
    49.0 & 307 & 0.37 & 10.0 & 0.7 & 94 \\
    77.5 & 389 & 0.35 & 10.0 & 0.5 & 95 \\    
    \bottomrule

    \end{tabular}
    \tablefoot{Inverse solutions for STEREO A, derived using the 1D SEP propagator model. For each species and energy (rigidity), we provide $\lambda_{\parallel}$, acceleration ($t_a$), and escape times ($t_e$) as well as the $R^2$ scores of the best fits.}
\label{tab:tab2}
\end{table}
%%%%end table%%%

\begin{figure}[h!]
\centering
\includegraphics[width=0.48\textwidth]{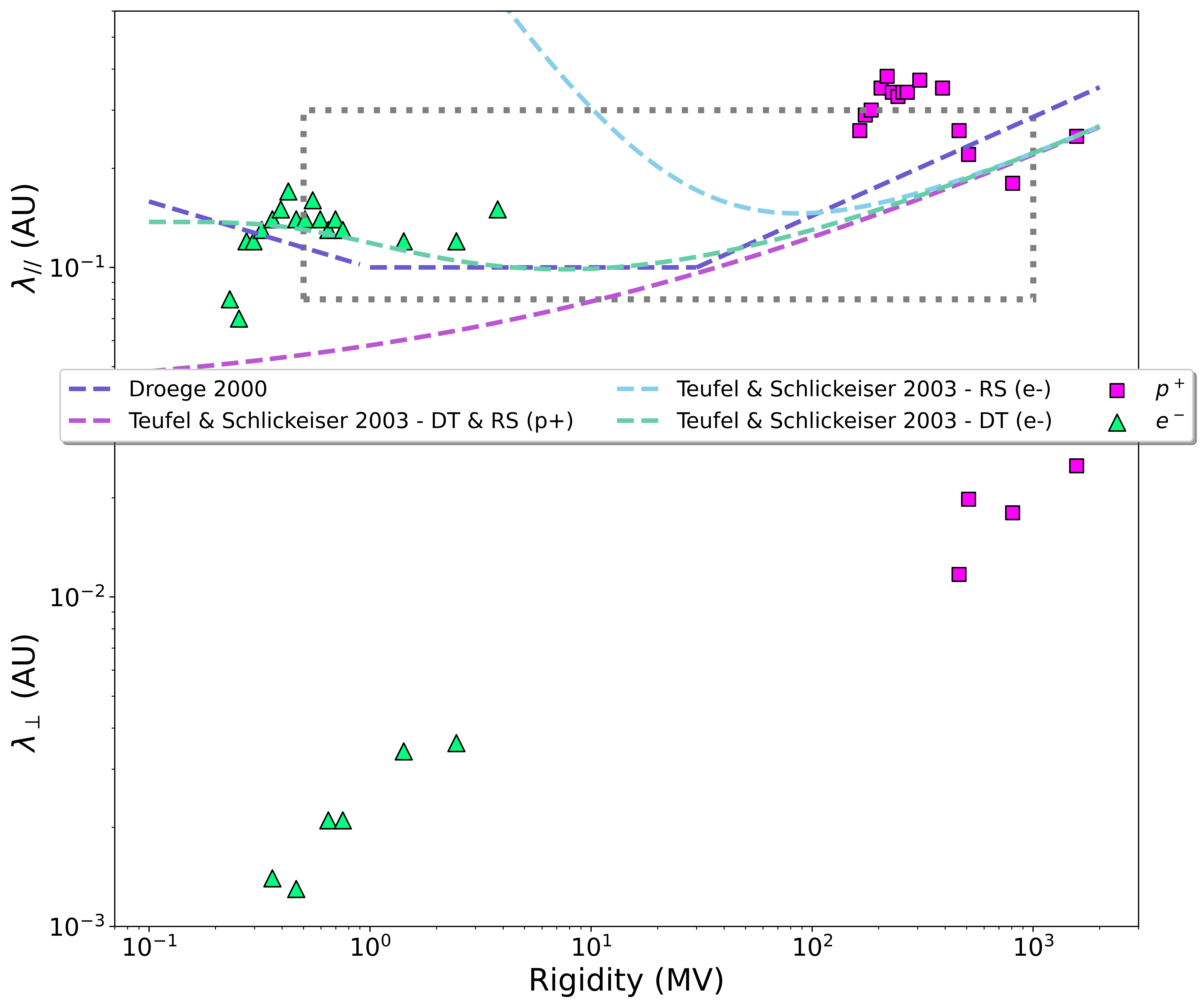}
\caption{Rigidity dependence of the parallel and perpendicular mean free path, as inferred by the inverse solutions listed in Tables \ref{tab:tab2} and \ref{tab:tab3}. The green triangles and magenta squares correspond to electrons and protons, respectively. The area denoted by the dashed gray line corresponds to the Palmer consensus. The other dashed curves depict the empirical relation suggested by \citet{Droge_2000} (purple), the analytical approximations of \citet{Teufel_2002, Teufel_2003} for electrons under damping turbulence (DT)  and random sweeping (RS) (green and cyan, respectively);  the corresponding relation for protons under both DT and RS is also depicted (violet).} 
\label{fig:mfp_rig_1d}
\end{figure}

We investigated the rigidity dependence of derived $\lambda_{\parallel}$ values, and show the  results compared with previous studies in Fig.~\ref{fig:mfp_rig_1d}. The best‑fit solutions indicate $\lambda_{\parallel}\sim$0.07--0.18~AU for electrons and 0.25--0.38~AU for protons, in violation of the Palmer consensus (0.08--0.30~AU) for high rigidity protons, as also found in previous studies \citep{bieber_1994,Engelbrecht_2022}. A further comparison was performed against the trends empirically identified by \citet{Droge_2000}, theoretically approximated by \citet{bieber_1994}, and extended by \citet{Teufel_2002,Teufel_2003} for SEP focused transport under dynamic turbulence. It is worth noting that many of the studies used for comparison rely on quasi-linear or related analytical descriptions. More recently, test-particle simulations in fully dynamical magnetic turbulence have been presented by \citet{2016ApJ...817..136H}, which provide an alternative and more self-consistent numerical treatment of energetic particle transport. While the present work focuses on analytical modeling, the results of such simulations offer a useful context for interpreting the trends discussed here. We find that protons in our sample follow the expected trends, with increasing $\lambda_{\parallel}$ for higher rigidities, in the range 180--400~MV. In contrast, low-rigidity electrons in the range 0.20--0.45~MV seem to follow a trend that is opposite to the expected trend, where $\lambda_{\parallel}$ increases for higher rigidities instead of decreasing (random sweeping) or being rather stable (damping turbulence). However, we note that for a fine-tuned parameterization of turbulence parameters, the DT approximation could better explain our results.  

By inspecting the acceleration and escape times derived for the electrons under the assumption of a time-dependent injection with a Reid-Axford profile, we find that $t_a$ increases from 0.5~h to 2~h for higher energies, while $t_e$ decreases within the same range (2~h to 0.5~h), with a few exceptions. This is consistent with physical expectations based on the possible operating acceleration mechanisms (e.g., DC flare acceleration in magnetic reconnection) that higher energies are accelerated for a longer time and escape quickly from the acceleration site. We  determined that the adopted type of injection profile can explain the derived solutions. However, inverse solutions for protons suggest too long acceleration times (5--12~h) so that a different type of time-dependent injection, for example simulating the CME-driven shock propagation, as in \citet{Jarry_2024}, could provide more physically explainable results regarding proton injection. 

\subsubsection{Anisotropy and pitch-angle diffusion}

\begin{figure}[h!]
\centering
\includegraphics[width=0.43\textwidth]{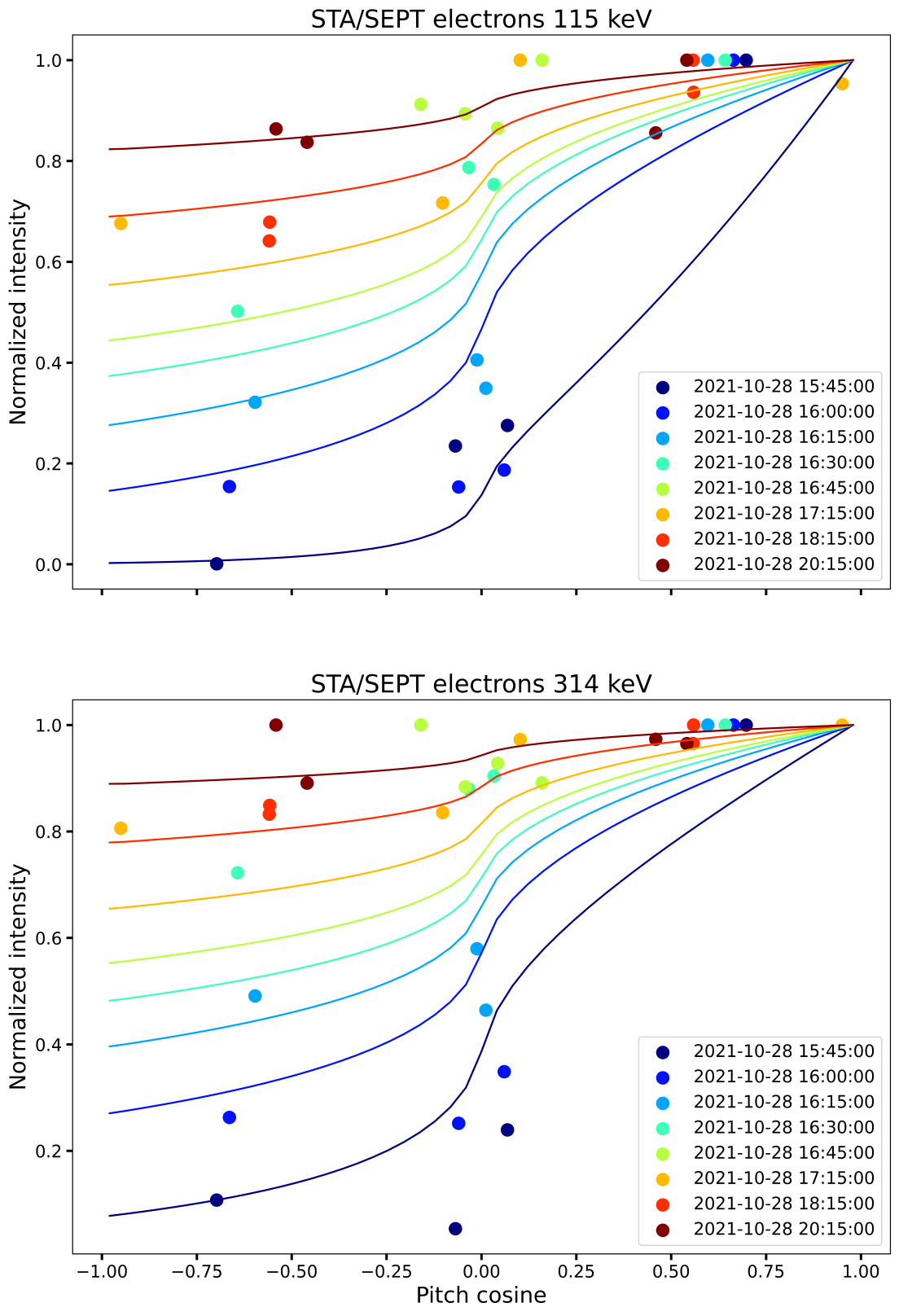}
\caption{Evolution of PAD for 115 and 314~keV electrons, from $\sim$0.5~h (dark blue) to 5~h (brown) after the injection. Actual measurements from the four STA/SEPT sectors are depicted as circles and the corresponding simulated PADs as solid lines with the same color. The color-coding of the successive timestamps is explained in the legend.} 
\label{fig:pad_115keV}
\end{figure}

Solutions for STA/SEPT electrons were further validated by comparing the simulated anisotropies to the observed properties; an example is presented in Fig. \ref{fig:e_314_1d} (bottom panel). We find that the simulated anisotropies capture the main variation in the actual profiles. 
Regarding the temporal evolution of anisotropy during the event, we plot simulated PADs for 115~keV and 314~keV electrons at successive time steps ranging from 0.5 to 5~hours after the injection in Fig.~\ref{fig:pad_115keV}. For the same time  steps, we overplot the normalized actual intensities recorded by the different STA/SEPT sectors at their corresponding (mean) pitch angles in each time step. We observe that the simulated PADs are generally in good agreement with the actual observations, showing highly anisotropic distributions at the start of the event, which gradually become isotropic after $\sim$3~hours from injection, due to the effect of small-scale magnetic turbulence that drives pitch-angle scattering.

%%%%%%%%%%%%%%%%%%%%%%%%%%%%%%%%%%%%%%%%%%%%%%%%%%%%%%%%%%%%%%%%%%%%%%%%%%%
%%%%%%%%%%%%%%%%%%%%%%%%%%%%%%%%%%%%%%%%%%%%%%%%%%%%%%%%%%%%%%%%%%%%%%%%%%%%%%

%%%%%%%%%%%%%%%%%%%%%%%%%%%%%%%%%%%%%%%%%%%%%%%%%%%%%%%%%%%%%%%%%%%%%%%%%%%
\begin{figure*}[h!]
\centering
\includegraphics[width=0.95\textwidth]{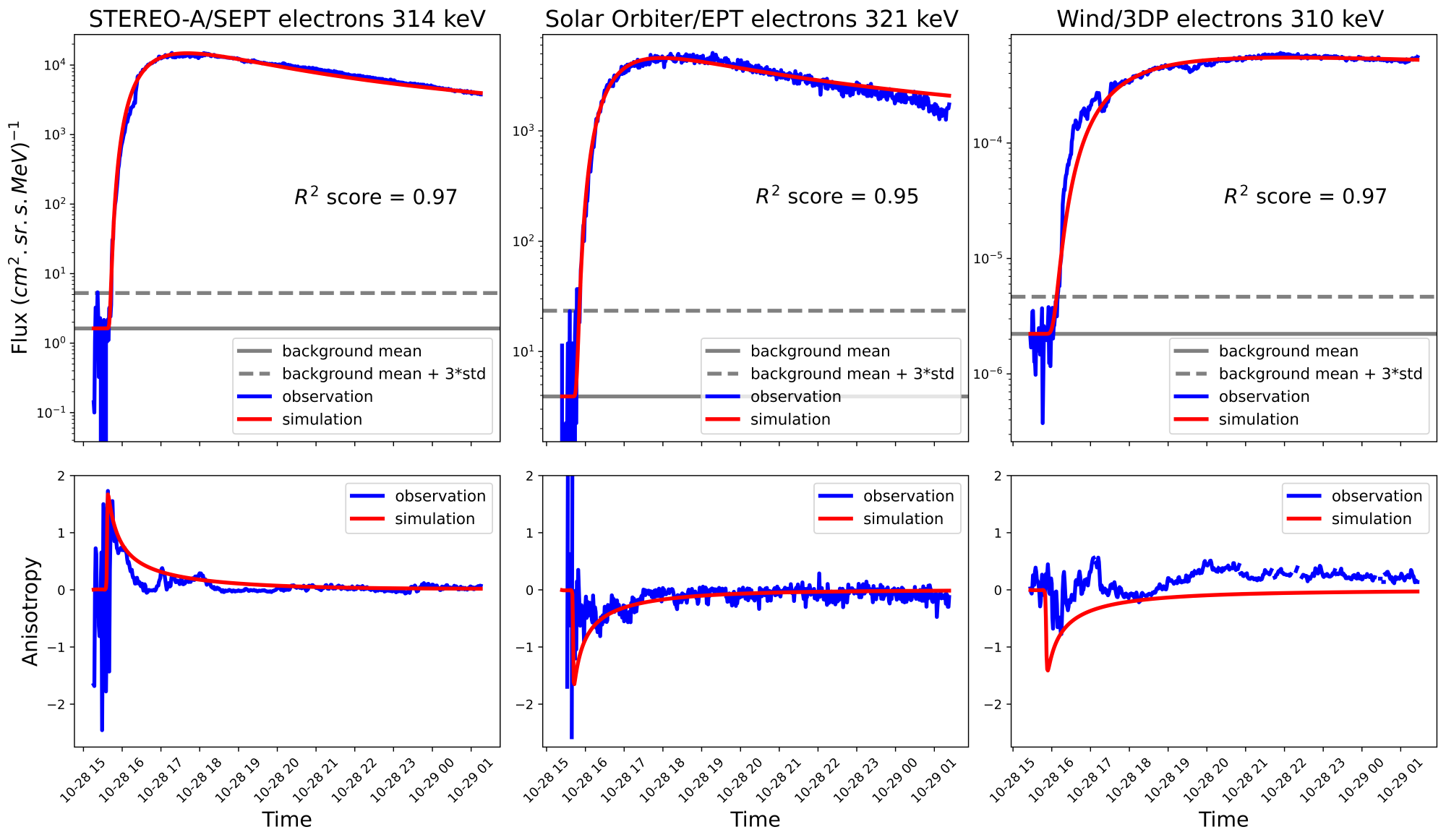}
\caption{Inverse solution to particle flux (top) and anisotropy (bottom) of three observers (left: STA/SEPT, middle: Solar Orbiter/EPT, right: Wind/3DP) for $\sim$314~keV electrons. The observed particle fluxes and anisotropies are depicted in blue and the simulated ones in red.  The  solid gray line in the flux profiles indicates the background level and the dashed line the background level plus 3$\sigma$. Simulated anisotropies have been reversed for Solar Orbiter and Wind, which were in an opposite polarity HMF to STEREO (see Fig.~\ref{fig:pads_314keV} in the Appendix).} 
\label{fig:2d_fit}
\end{figure*}
%%%%%%%%%%%%%%%%%%%%%%%%%%%%%%%%%%%%%%%%%%%%%%%%%%%%%%%%%%%%%%%%%%%%%%%%%%%%%%

%%%%%%%%%%%%%%%%%%%%%%%%%%%%%%%%%%%%%%%%%%%%%%%%%%%%%%%%%%%%%%%%%%%%%%%%%%%
\begin{figure}[h!]
\centering
\includegraphics[width=0.44\textwidth]{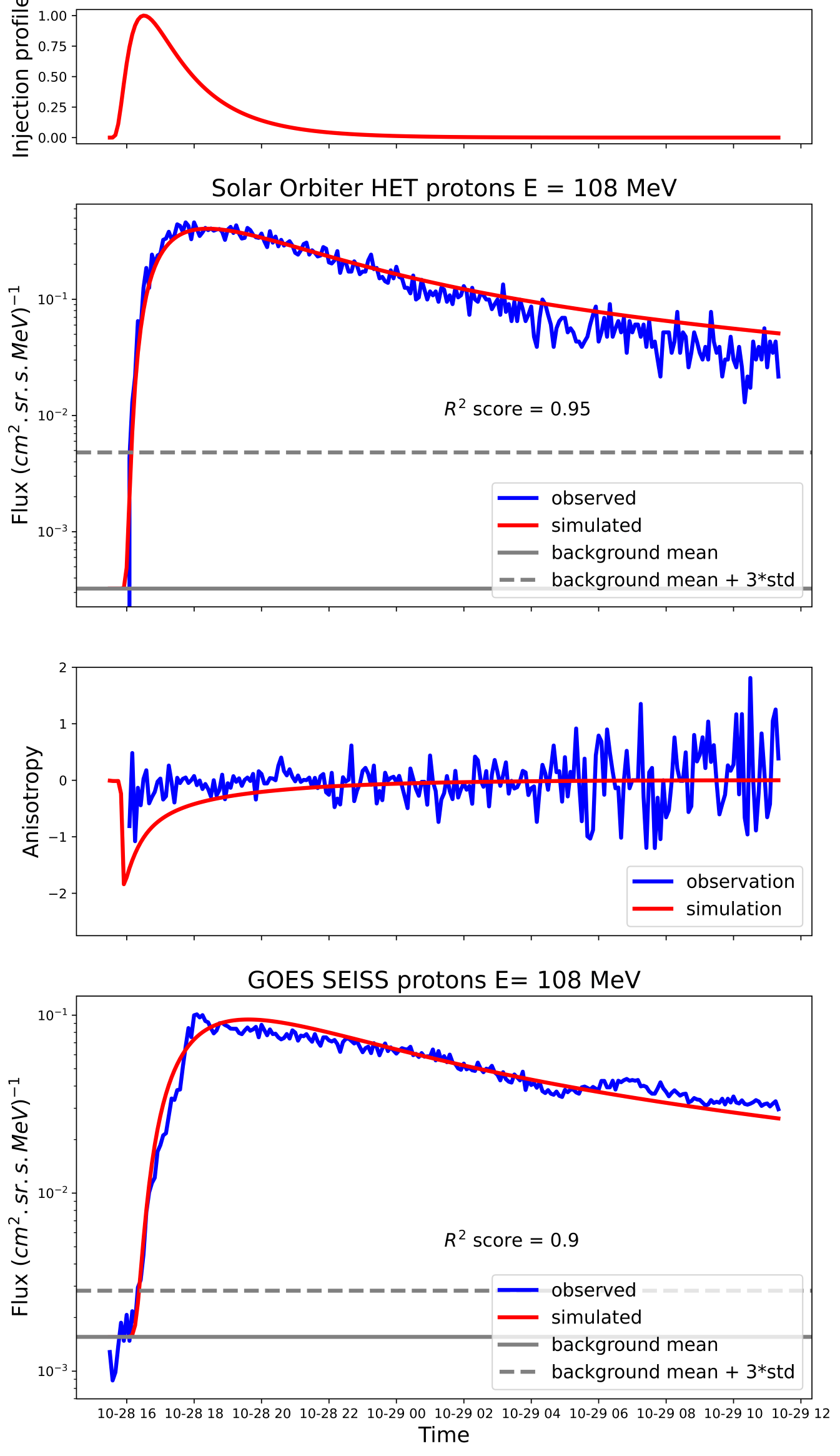}
\caption{Inverse solution to 108 MeV protons at Solar Orbiter and Earth. Shown are (from top to bottom)  the injection profile, particle flux and anisotropy at Solar Orbiter, and particle flux at GOES. The observed particle fluxes and anisotropies are depicted in blue, and the simulated ones in red.  The solid gray  line in the flux profiles indicates the background level and the dashed line the background level plus 3$\sigma$.} 
\label{fig:proton_fit_108}
\end{figure}
%%%%%%%%%%%%%%%%%%%%%%%%%%%%%%%%%%%%%%%%%%%%%%%%%%%%%%%%%%%%%%%%%%%%%%%%%%%%%%

\subsection{Multiple observers} \label{sec:shock_reco}

\subsubsection{Perpendicular diffusion and injection broadness}

\begin{figure}[h!]
\centering
\includegraphics[width=0.45\textwidth]{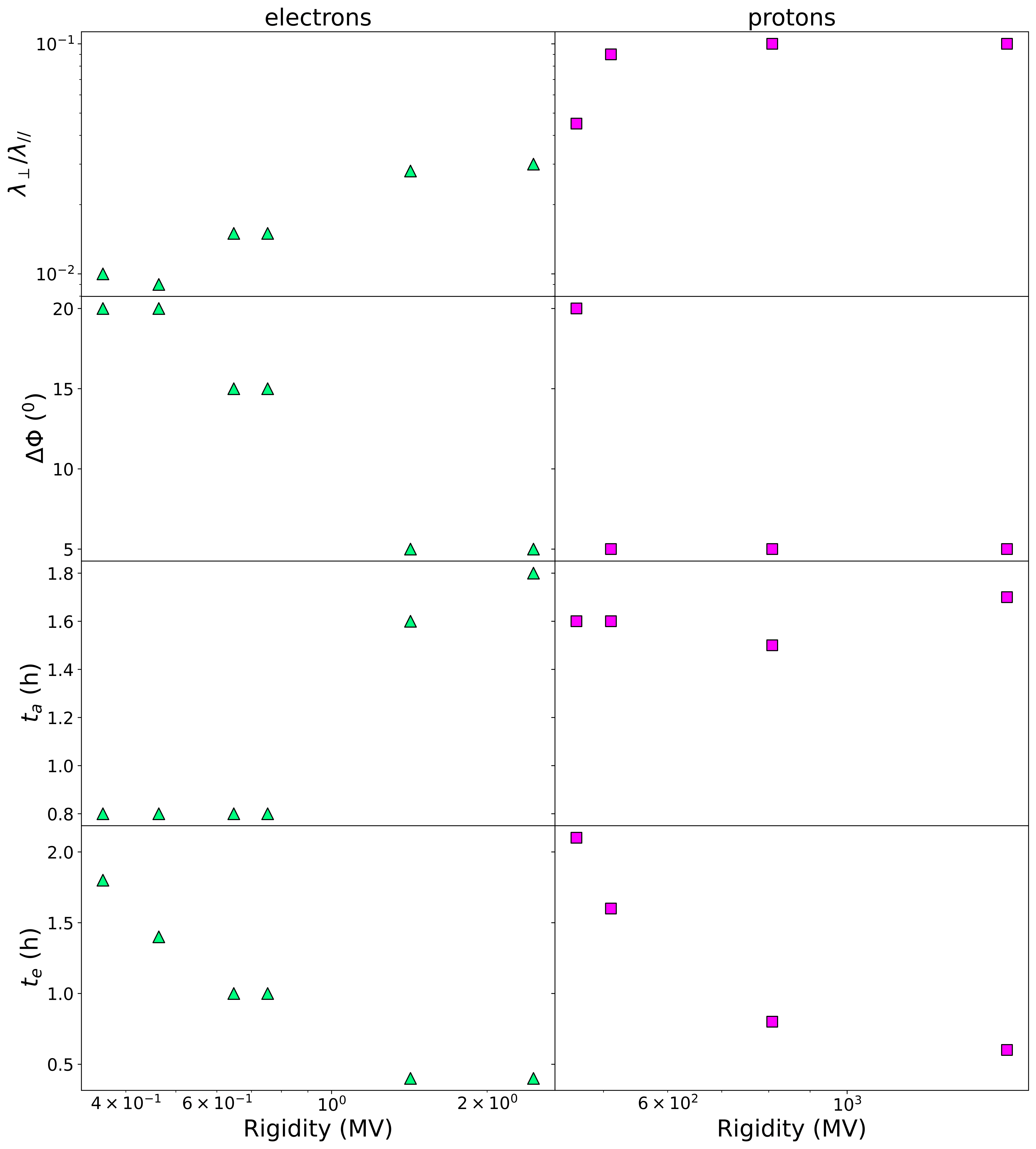}
\caption{Rigidity dependence of 2D transport parameters. Shown from top to bottom are the inverse solutions vs. rigidity, for the ratio $\lambda_{\perp} / \lambda_{\parallel}$, injection broadness $\Delta\Phi$, acceleration time $t_a$, and escape time $t_e$.  The green triangles correspond to electrons and the magenta squares  to high energy protons. } 
\label{fig:rigidity_2d}
\end{figure}

We found inverse solutions of the 2D simulations on all available observers, to constrain the particles’ $\lambda_{\perp}$, as well as the size of the injection source. The optimal parameters recovered and the corresponding fit scores (individual and average) for protons (108, 130, 300, and 896~MeV) and electrons (115, 180, 314, and 400~keV; and 1.0 and 2.5~MeV) are listed in Table~\ref{tab:tab3}. We provide an example best-fit solution for electrons at 314~keV in Fig.~\ref{fig:2d_fit}, comparing particle intensities and anisotropies at all three observers,  STA, SolO, and Earth.  Similarly, in Fig.~\ref{fig:proton_fit_108} we display  the comparison of the best-fit solution to 108 MeV protons detected by SolO/HET and GOES/SEISS; the anisotropies are only available for the former. In Fig.~\ref{fig:mfp_rig_1d} and in Fig.~\ref{fig:rigidity_2d} (top panel) we  plot $\lambda_{\perp}$ and $\lambda_{\perp}/\lambda_{\parallel}$, respectively, against particle rigidity. We find $\lambda_{\perp}\sim$~0.0013--0.0036~AU for electrons and $\lambda_{\perp}\sim$~0.012--0.025~AU for protons. The results suggest that appreciable cross‑field diffusion is required to explain the broad longitudinal spread, with $\lambda_{\perp}/\lambda_{\parallel} \simeq$~5--10\% for protons and $\simeq$~1--3\% for electrons. A slight increase within those limits is observed, with increasing particle rigidity, for both species, but more pronounced for the electrons in our sample, due to wider rigidity coverage.

%%%%%%%%%%%%%%%%%%%%%%%%%%%%%%%%%%%%%%%%%%%%%%%%%%%%%%%%%%%%%%%%%%%%%%%%%%%
%%%%%%%%%%%%%%%%%%%%%%%%%%%%%%%%%%%%%%%%%%%%%%%%%%%%%%%%%%%%%%%%%%%%%%%%%%%%%%

The inverse solutions in Table~\ref{tab:tab3} suggest that the size of the injection source, $\Delta\Phi$, is within 5--20$^{\circ}$ for both protons and electrons. The injection size is found to decrease with increasing rigidity for both species, suggesting that higher-rigidity particles are injected into  narrower regions (Fig.~\ref{fig:rigidity_2d}, lower panel). 

%%%%%Begin table%%%%%%%
\begin{table*}[!ht]
\footnotesize
     \centering
     \caption{Inverse solutions for multiple observers, using the 1D and 2D SEP propagator models.}
     \label{tab:tab3}
    \begin{tabular}{ ccccccccccccc}
    \toprule
    \textbf{Energy} & \textbf{Rigidity} & \textbf{$\lambda_{\parallel}$} & \textbf{$\lambda_{\perp}$} & \textbf{$ \lambda_{\perp} / \lambda_{\parallel}$} & \textbf{Injection} & \textbf{Injection} & \textbf{$t_a$} & \textbf{$t_e$} & \textbf{$R_{STA}^2$} & \textbf{$R_{SolO}^2$} & \textbf{$R_{Earth}^2$} & \textbf{$R_{mean}^2$} \\
    \textbf{(MeV)} & \textbf{(MV)} & \textbf{(AU)} & \textbf{(AU)} & \textbf{\%} & \textbf{source ($^{\circ}$)} & \textbf{time (UT)} & \textbf{(hours)} & \textbf{(hours)} & \% & \% & \% & \% \\
    \hline
    \multicolumn{13}{c}{\textbf{Electrons}} \\
    \hline
    0.115 & 0.361 & 0.14 & 0.0014 & 1 & 20 & 2021-10-28 15:17 & 0.8 & 1.8 & 98 & 96 & 92 & 95 \\
    0.180 & 0.463 & 0.14 & 0.0013 & 0.9 & 20 & 2021-10-28 15:20 & 0.8 & 1.4 & 98 & 95 & 91 & 95 \\
    0.314 & 0.647 & 0.13 & 0.0021 & 1.5 & 15 & 2021-10-28 15:24 & 0.8 & 1.0 & 97 & 95 & 97 & 96 \\
    0.400 & 0.752 & 0.13 & 0.0021 & 1.5 & 15 & 2021-10-28 15:23 & 0.8 & 1.0 & 97 & 92 & 88 & 92 \\
    1.0 & 1.4 & 0.12 & 0.0034 & 2.8 & 5 & 2021-10-28 15:19 & 1.6 & 0.4 & 93 & 96 & & 95 \\
    2.5 & 2.46 & 0.12 & 0.0036 & 3 & 5 & 2021-10-28 15:18 & 1.8 & 0.4 & 95 & 98 & 91 & 95 \\
    \hline \hline
    \multicolumn{13}{c}{\textbf{Protons}} \\
    \hline
    108 & 463 & 0.26 & 0.0117 & 4.5 & 20 & 2021-10-28 15:33 & 1.6 & 2.1 & & 95 & 90 & 92 \\
    130 & 511 & 0.22 & 0.0198 & 9 & 5 & 2021-10-28 15:23 & 1.6 & 1.6 & & 99 & 83  & 91 \\
    300 & 808 & 0.18 & 0.018 & 10 & 5 & 2021-10-28 15:23 & 1.5 & 0.8 & & 98 & & 98 \\
    896 & 1576 & 0.25 & 0.025 & 10 & 5 & 2021-10-28 15:28 & 1.7 & 0.6 & & 96 & & 96 \\
    \bottomrule

    \end{tabular}
\tablefoot{For each species and energy (rigidity) we provide $\lambda_{\parallel}$, $\lambda_{\perp}$, and their ratio $\lambda_{\perp} / \lambda_{\parallel}$; the size of injection source, injection time, acceleration time ($t_a$), and escape time ($t_e$); and the $R^2$ scores of the best fits.}
\end{table*}
%%%%end table%%%

\subsubsection{Evidence for unique solutions}

\begin{figure}[h!]
\centering
\includegraphics[width=0.44\textwidth]{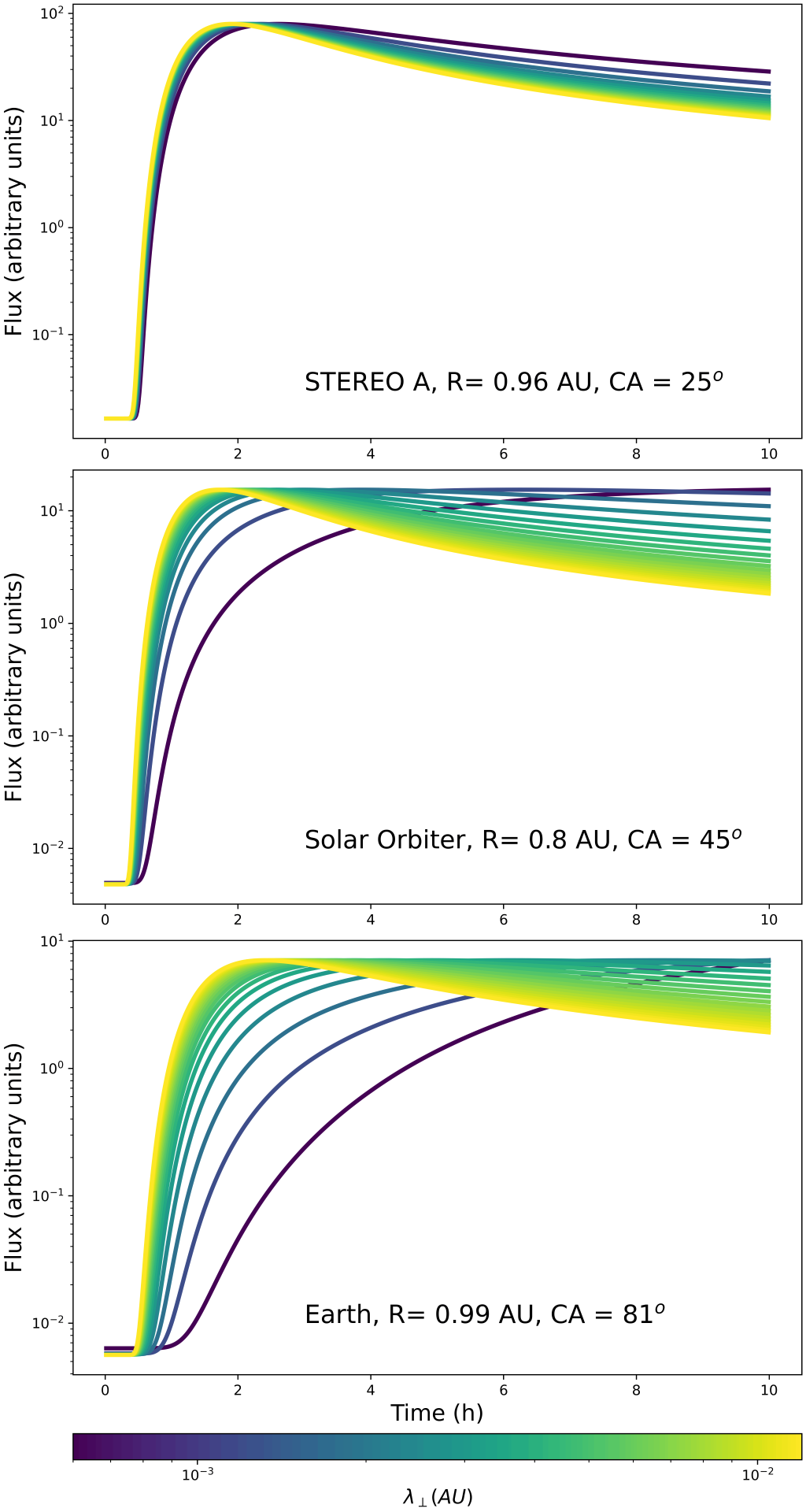}
\caption{Simulated flux profiles for $\sim$2.5~MeV electrons at STA (top), Solar Orbiter (middle), and Earth (bottom) under variation of the perpendicular mean free path $\lambda_{\perp}$ with all other parameters set to the best-fit solutions. } 
\label{fig:var_mfp}
\end{figure}

\begin{figure}[h!]
\centering
\includegraphics[width=0.44\textwidth]{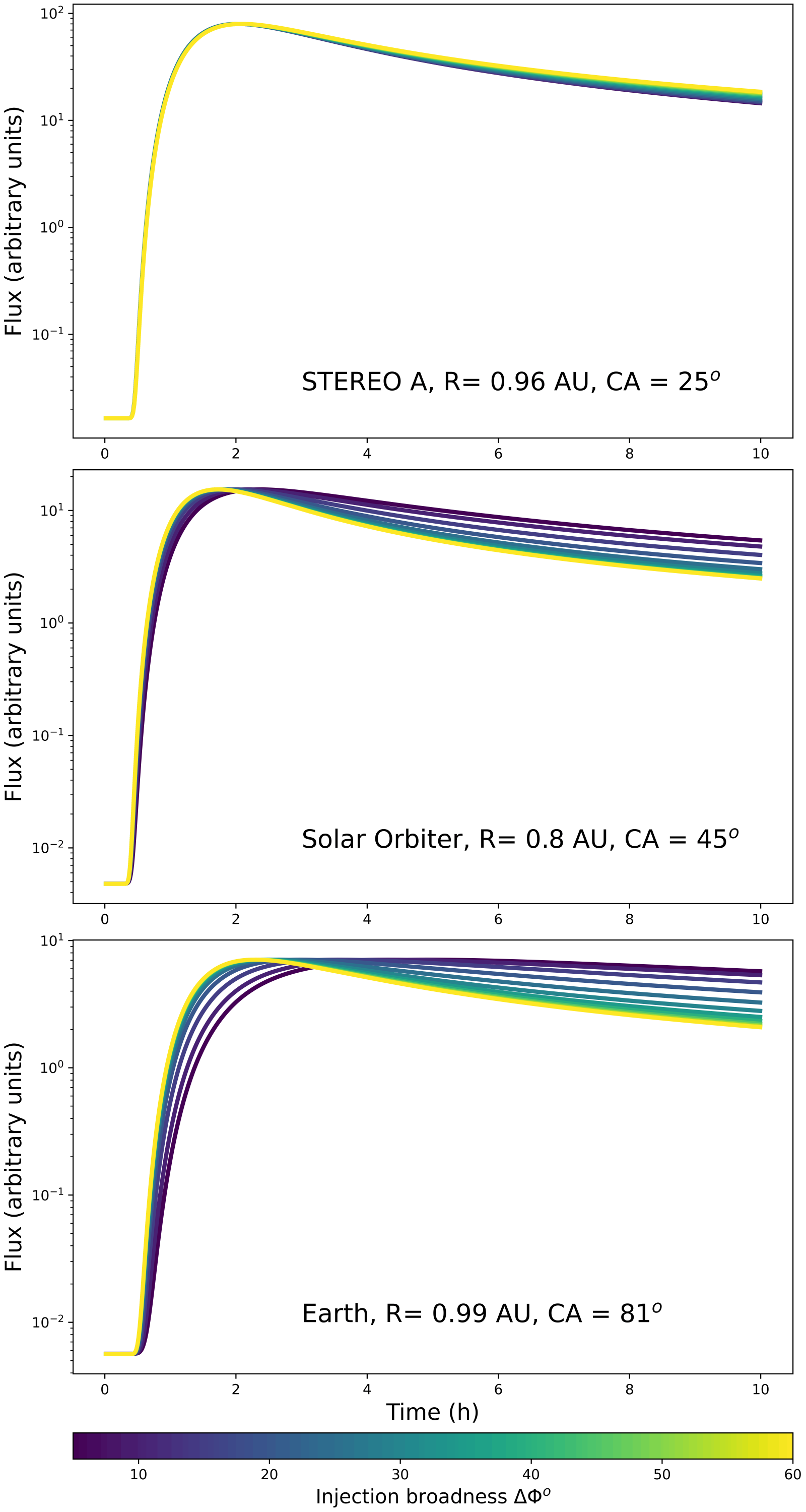}
\caption{Simulated flux profiles for $\sim$2.5~MeV electrons at STA (top), Solar Orbiter (middle), and Earth (bottom) under variation of the size of injection source, with all other parameters set to the best-fit solutions. } 
\label{fig:var_injection}
\end{figure}

We investigated the dependence of the solutions at multiple observers on each of the two free parameters ($\lambda_{\perp}$, injection size) to check whether there is the possibility of a solution space where the same result could be achieved  either with a narrow injection and strong perpendicular diffusion or with a wide injection and weak perpendicular diffusion. Our results indicate that this is not possible, as the observed time profiles can only be reproduced with a narrow injection and significant cross-field diffusion. A wide source with any level of perpendicular diffusion used in our setting fails to reproduce the observations. We note here our modeling limitation in the use of a single form of injection profile, meaning that this observation applies to simulations under the specific injection function, and we cannot testify as to how results would be affected by the choice of a different type of injection, as in \citet{Jarry_2024}, among others. Resolving this limitation would require more flexibility in the modeling to allow for different injection profiles. Nonetheless, for the 2.5~MeV electrons, we stabilized all the optimal parameters (see Table~\ref{tab:tab3}) and varied $\lambda_{\perp}$ (Fig.~\ref{fig:var_mfp}) and the size of injection source (Fig.~\ref{fig:var_injection}). The plots indicate that the solution is highly sensitive to changes in the perpendicular mean free path, especially for the less connected observers which seem to constrain the solution. Results indicate that the observed time profiles at the different observers can only be obtained with a narrow source and considerable ($\sim$3\%) cross-field diffusion, and cannot be reproduced by a wide source and insignificant ($<$1\%) cross-field diffusion. These results support the uniqueness of the solutions, corroborated also by the evaluation scores in the different ranges of parameter values, within the tested parameter space.  

\subsubsection{Particle injection times}
Particle injection times were estimated by the temporal alignment of simulated and observed time profiles for the best-fit 2D simulations. The inferred times, corrected for the light travel time to the spacecraft ($\sim$8~min), are presented in Table \ref{tab:tab3}. We find the average injection time of protons is around 15:27 $\pm$~5~min and that of electrons is around 15:20 $\pm$~3~min. These are in good agreement with the GOES SXR flare start at 15:17 and the first-order calculated onset of the CME at 15:25 \citep{Papaioannou2022}. 
The estimated release time of protons is also close to the start of the first group of DH type III emission (at frequencies below 20~MHz), as well as the onset of the type II radio burst at 15:28 UT. For the electrons, the estimated release time aligns well with the early phase of radio emission in high frequencies (above 100 ~MHz), where spike bursts and fast reverse drifts along with the rise of hard X-ray (HXR) emission signify early episodes of electron acceleration \citep{Klein_2022}.

\subsubsection{Electron onset delay versus connectivity}
We examined the efficiency of perpendicular diffusion with respect to focused transport for this event, and tested whether the 2D SEP-propagator model reproduces the observables. To do so, we inspected the event onset delay at the different observers in relation to their magnetic connectivity to the eruption site. Event onsets at STA, SolO, and the Earth were defined by the 3$\sigma$ method applied to intensity-time profiles, as explained in Section \ref{sec:3}, for the examined particle energies listed in Table \ref{tab:tab3}. Corresponding onsets were defined for the three observers in the best-fit simulated intensity time profiles, as the times when intensity exceeds $1/1000$ of the maximum intensity across observers, similarly to \citet{Strauss_2017, Strauss_2023}. Real and simulated onset delays were calculated by the time difference between the parent SXR flare start at 15:17 UT and the respective real and simulated event onsets. In an attempt of a crude analytical approximation of onset delay versus connectivity, we adopted the approach of \citet{Strauss_2023}, assuming scatter-free propagation of SEPs to the magnetically connected observer and perpendicular diffusion due to magnetic turbulence otherwise. Under this approach, the onset delay for the optimal magnetic connection would be $\tau_0 = \Delta s_{\parallel}/v$, with $\Delta s_{\parallel}$ the nominal Parker spiral length to the observer (according to their radial distance and solar wind speed) and $v$ the particle velocity. For a disconnected observer at magnetic connection angle $CA$ and an isotropic distribution of particles, this is augmented by the diffusive propagation time so that the onset delay can be estimated as 

\begin{figure}[h!]
\centering
\includegraphics[width=0.49\textwidth]{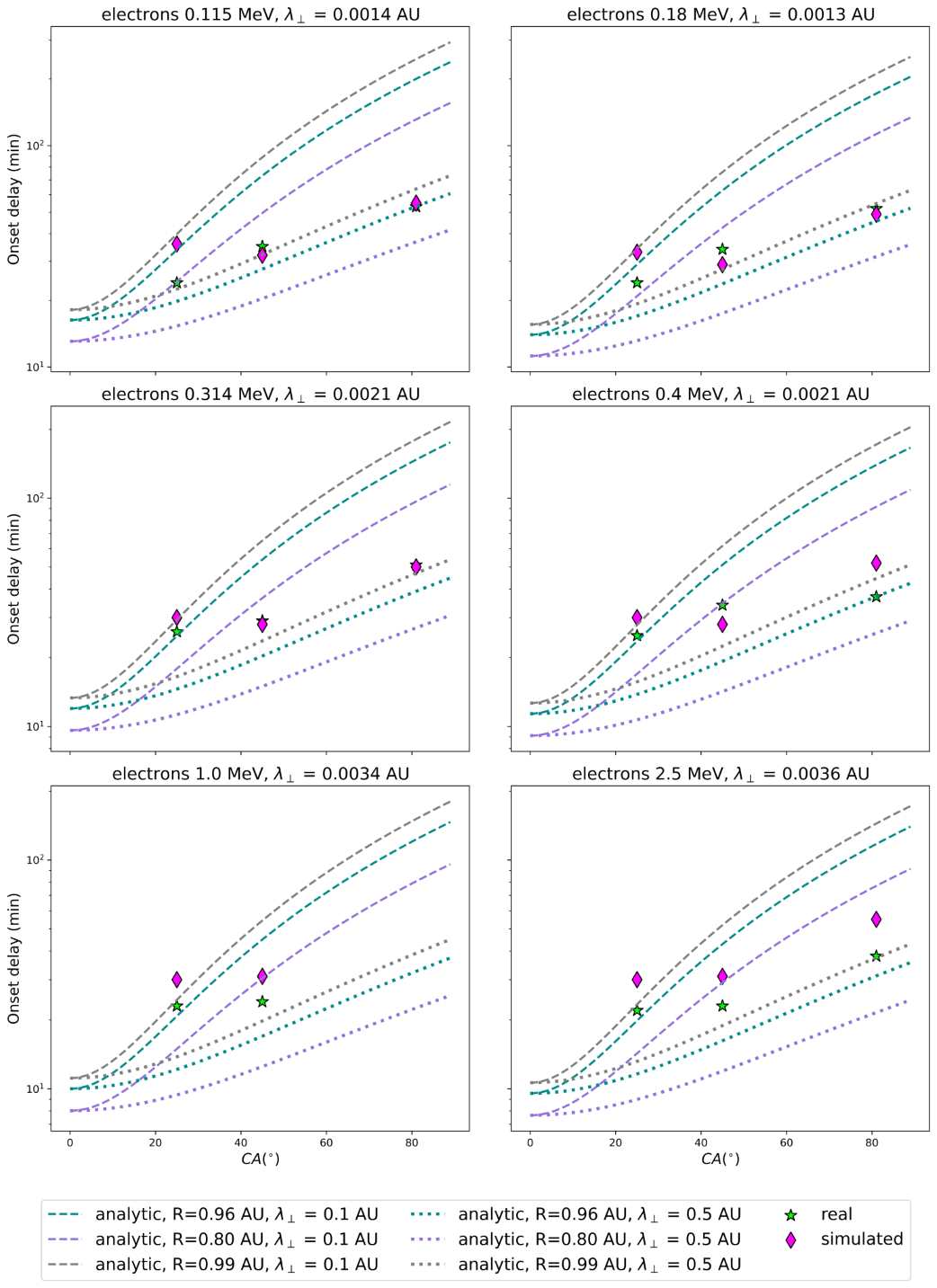}
\caption{Onset delay vs. connection angle for 0.115--2.5~MeV electrons studied in the 2D transport analysis. The real and simulated onsets are depicted as green stars and magenta diamonds, respectively. The dashed and dotted lines show analytical approximations at each observer for $\lambda_{\perp}$ equal to 0.1 and 0.5 AU, respectively.} 
\label{fig:onset_delay}
\end{figure}

\begin{equation}
    \tau = \frac{\Delta s_{\parallel}}{v} \left [  1 + \frac{\Delta s_{\parallel}}{2 \lambda_{\perp}} (CA)^2 \right ].
    \label{eq:analytic}
\end{equation}

The real and simulated onset delays with respect to connection angle are displayed in Figure \ref{fig:onset_delay}; the analytic approximations applying Equation \ref{eq:analytic} to the nominal spiral length at each observer and for different values of $\lambda_{\perp}$ (0.1 and 0.5 AU) are overplotted.  We find that the real onset delays are very close to the observed ones, suggesting that the efficiency of perpendicular diffusion implemented in the model can reproduce  the observations quite well. We note that the model suggests a slightly earlier arrival of lower-energy electrons in our sample at SolO (r = 0.80 AU, CA = $45^{\circ}$)  compared to STA (r = 0.96 AU, CA = $25^{\circ}$). This implies that perpendicular diffusion at this range of energies is a bit more effective in the model than in reality, and that a modified perpendicular diffusion coefficient that is less effective at lower energies could better fit the observations. However, also considering  the uncertainty induced mainly in the real onset determination depending on pre-event background, the differences are small and on the order of the accuracy of the onset determination ($\sim 1$ minute). Real and simulated onset delays also fall within analytical estimates for $\lambda_{\perp}$ = 0.1, 0.5 AU, as in the study of multiple events conducted by \citet{Strauss_2023}. We note that these values are comparable to the simplified approximation of $\Delta S_{\perp} \simeq r \times CA$ for STA (0.42 AU) and SolO (0.63 AU).

\section{Discussion and conclusions}
\label{sec:5}
A comprehensive case study on particle transport during the relativistic SEP event on 28 October 2021 was conducted, with the use of numerical simulations and observational data, covering a wide range of proton and electron rigidities. Inverse solutions were recovered for multiple observers at different vantage points from the eruption site to model the transport of particles along and across the HMF during this powerful event. High particle anisotropies recorded in magnetically well-connected vantage points characterize the event up to $\sim$3~hours after particle release. 

Focused transport is found to be the dominant component of particle motion, under the combined effects of pitch-angle scattering and cross-field diffusion. The observed rigidity dependence of the parallel mean free path, $\lambda_{\parallel}$, is consistent with transport models that incorporate a dynamical turbulence description of the temporal correlation of magnetic field fluctuations, rather than a purely magnetostatic treatment \citep{Teufel_2003,2006ApJ...642..230S,Lang_2024}. Such models account for the  dynamic time-dependent nature of the turbulence that governs pitch-angle scattering, without implying any specific turbulence generation mechanism. Considerable cross-field diffusion is found necessary to explain the observed profiles at multiple longitudinally widespread observers, as supported also by previous studies on widespread events \citep[e.g.,][]{Dresing_2014}. The level of perpendicular-to-parallel diffusion seems to increase for higher rigidities, with protons experiencing stronger (5-10\%) perpendicular-to-parallel diffusion than electrons (1-3\%), possibly due to higher rigidity and gyro-radius.  

The estimated size of the injection region is found to be small, not exceeding 20$^{\circ}$ across species and rigidities, and converging to about 5$^{\circ}$ for both species at higher rigidities. This suggests that the injection region for high-energy particles during this event is highly localized. In contrast, \citet{Koulou2024} adopt an extended ($60^{\circ} \times 60^{\circ}$) source scenario, where the acceleration of protons up to 1~GeV is attributed to the fast and wide CME-driven shock. In their simulations, proton energies in the range 50--1000~MeV were modeled and compared with high-energy observations from GOES and other spacecraft. Their test-particle code \citep{2005A&A...436.1103D}, previously applied to heliospheric particle transport, assumes a Parker spiral magnetic field and incorporates gradient and curvature drifts, including drift effects along the HCS. Particle transport was modeled using prescribed pitch-angle scattering characterized by a parallel mean free path of 0.3 AU, while cross-field motion was treated separately through an explicit model of perpendicular diffusion or magnetic field structure. In this work, particles at PSP, STA, and SolO are explained by the shock alone, since these observers are (gradually) connected to strong and moderate shock regions. However, curvature and gradient drifts in the presence of the HCS are essential to partially explain high-energy proton detections at Earth, Mars, and BepiColombo, which are connected to weak shock regions. These modeling constraints indicate that while extended source scenarios can reproduce certain aspects of the data, they do not capture the full range of possible transport effects.

On the other hand, the localized injection of relativistic particles, combined with their longer acceleration times, as suggested by our results (Table \ref{tab:tab3}), appears partly consistent with some aspects of the trapping scenario discussed by \citet{Klein_2022}. These authors, based on a detailed joint analysis of radio, HXR, and microwave observations, proposed that relativistic particles were accelerated by the flare in the low corona and subsequently confined within extended magnetic structures beneath the rising CME. As the CME expanded, reconnection between the closed flux rope and the surrounding open magnetic field lines allowed these trapped particles to escape into interplanetary space. While our findings agree that the flare–CME system plays a central role in particle acceleration and release, the spatial characteristics of the injection region appear notably different. In contrast to the extended coronal source region inferred by \citet{Klein_2022}, our modeling indicates that a narrow well-defined injection region ($\leq$ 20$^{\circ}$) can account for the multi-spacecraft observations without requiring a large-scale coronal release. This suggests that the broad heliolongitude distribution of observed particles may result primarily from interplanetary transport effects rather than from a widespread coronal acceleration region.

Our analysis suggests that, under the adopted assumptions, the transport solutions across widely separated observers are unique; a given set of intensity profiles can be reproduced either by a narrow source with significant perpendicular diffusion or by an extended source with negligible diffusion, but not by both simultaneously. In the case of the GLE73 event, only a (relatively) narrow injection source with considerable cross-field diffusion could reproduce the observables. Additional processes may also contribute, including field-line meandering \citep{laitinen2016}, the evolution of interplanetary shocks, and multiple injections spanning a broad longitudinal sector \citep{2023A&A...674A.105D}, none of which are directly considered in our modeling framework. With regard, for example, to the effect of injection function selection on model results, the work in \citet{Jarry_2024} shows how model outputs change under this selection;  a propagating injection is the proper choice to explain observations in their case study. We note that the selected Reid-Axford injection function in our case seems to work better for the high-energy particles (electrons $\geq 110$ keV, protons $> 100$ MeV), whereas it implies unphysically long acceleration times for lower-energy protons in our sample. In this case, different functional forms of particle injection could reproduce the observables at this proton energy range offering at the same time a physical interpretation, as in the propagating injection using the modeled CME-driven shock characteristics applied in \citet{Jarry_2024}. 

Further, the fact that SEPs were observed by spacecraft located in regions of opposite HMF polarity strongly suggests that the particles propagated through the HCS in order to reach all observers, in agreement with \cite{Koulou2024} and \cite{2025ApJ...991..104W}. In addition, the injection source was in the southern hemisphere (S26 in Stonyhurst coordinates), while the three observers were right above the equatorial plane (latitudes of $\sim$2--7$^{\circ}$), although magnetically connected to the southern hemisphere according to the PFSS model (see Appendix \ref{appenA}). This brings out the limitation of our modeling approach to the equatorial plane in 2D with a uniform polarity magnetic field. As previous studies have shown, longitudinal and latitudinal spreads of SEPs can be explained by particle drifts in the presence of the HCS \citep[e.g.,][]{Battarbee_2018, dalla_2020}. However, these processes have not yet been modeled under the combined effect of cross-field diffusion due to magnetic turbulence. As the 2D SEP-propagator model can be naturally extended to 3D, the inclusion of HCS drifts could be a next step to expand our present analysis. 

Our results therefore indicate that a combination of localized injection and cross-field transport processes—including diffusion, field line meandering, and drifts along the HCS—offers a plausible explanation for the wide longitudinal spread observed during the relativistic SEP event of 28 October 2021. It is worth noting that GLE73 is a rather peculiar event. It was characterized by an Fe/O abundance ratio of 0.39 that was essentially invariant across different longitudes and radial distances \citep{2025ApJ...978L..35C}. This points to a hybrid origin of the high-energy particles, and rules out the possibility of SEP transport being accountable for the invariance of the Fe/O.  

A major challenge to extending the multi-observer analysis to the higher proton energies is the lack of common high energies recorded in the various spacecraft of the current fleet of heliospheric missions. Penetrating protons at very high energies constitute a major radiation hazard for astronauts in space. Therefore, the design of future science missions should consider filling this gap so that multi-spacecraft observations at the very high proton energy range can be used to model and validate particle transport.  

%------------------------------------------------------------------------

\begin{acknowledgements}
This research received funding from the European Union’s Horizon Europe programme under grant agreement No 101135044 (SPEARHEAD) [\url{https://spearhead-he.eu/}]. Views and opinions expressed are, however, those of the author(s) only and do not necessarily reflect those of the European Union or the European Health and Digital Executive Agency (HaDEA). Neither the European Union nor the granting authority can be held responsible for them. A.K. acknowledges financial support from NASA’s HGIO grant 80NSSC24K0555 and from NASA’s LWS grant 80NSSC25K0130.

\end{acknowledgements}

%------------------------------------------------------------------------

%------------------------------------------------------------------------
\bibliographystyle{aa}
\bibliography{references}

\begin{appendix} 
\onecolumn
\section{Magnetic field and plasma measurements at STA, SolO, and the Earth} \label{appenA}
\begin{figure}[h!]
\centering
\includegraphics[width=0.44\textwidth]{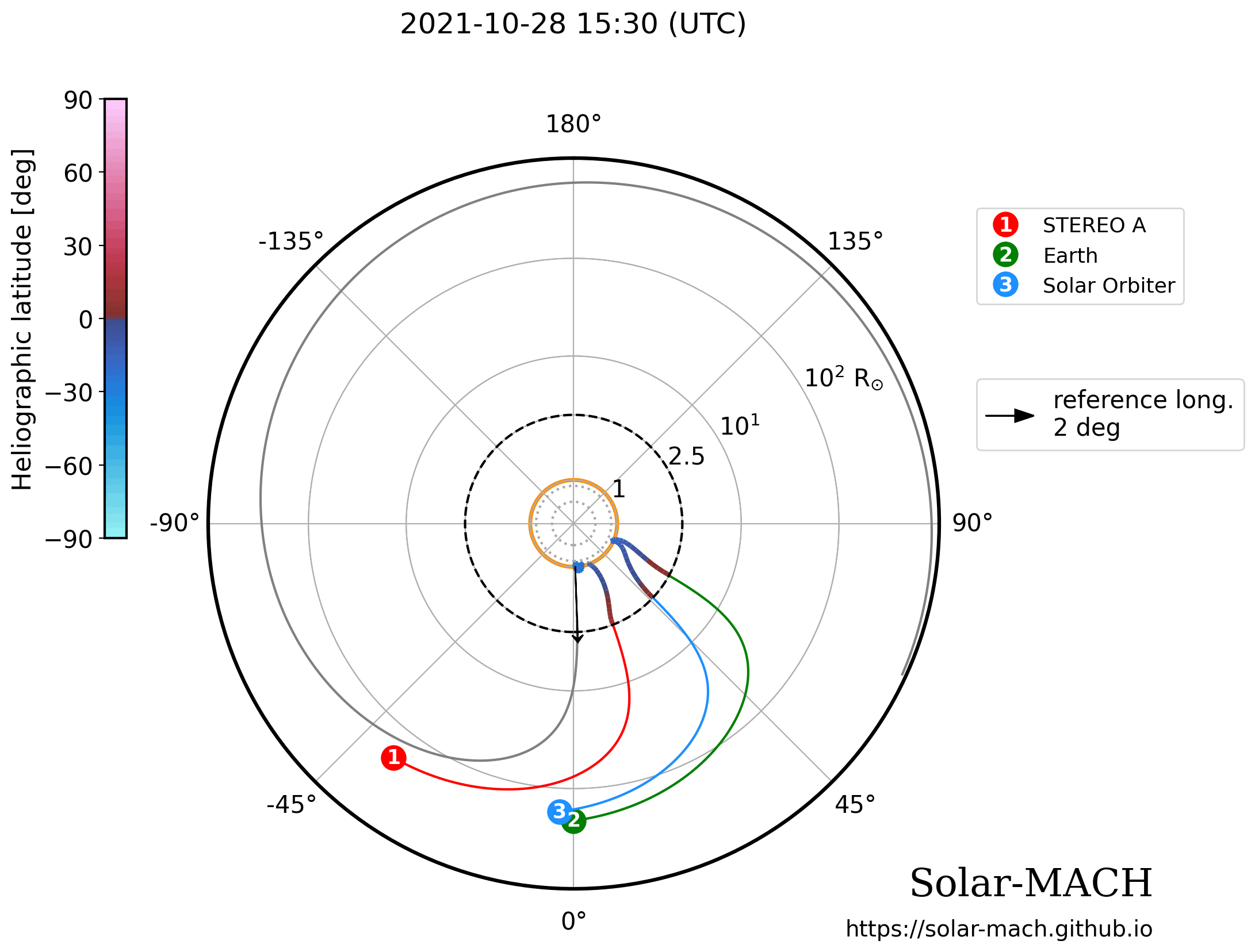}
\caption{Potential field source surface approximation for the event. } 
\label{fig:pfss}
\end{figure}

The potential field source surface (PFSS) approximation provides a global three-dimensional view of the coronal magnetic field topology at the time of the event. In Figure \ref{fig:pfss} we provide this view, generated with the Solar-MACH tool.\footnote{\url{https://solar-mach-pfss.streamlit.app/}} We note that magnetic field lines reaching STEREO-A, Solar Orbiter, and the Earth originate in the southern hemisphere of the sun, where the parent solar eruption occurred (W02S26).  

% %oooooooooooooooooooooooooooooooooooooooooooooooooooooooooooooooooooooooo
% %oooooooooooooooooooooooooooooooooooooooooooooooooooooooooooooooooooooooo

We inspected magnetic field and plasma observations at STA, SolO, and the Earth to ensure that no major disturbances are present in the HMF and the solar wind. This affirms the validity of the adopted form of the HMF as a Parker spiral on average, with small-scale turbulence. 

%oooooooooooooooooooooooooooooooooooooooooooooooooooooooooooooooooooooooo
\begin{figure*}[h!]
    \centering
        \includegraphics[width=0.31\textwidth]{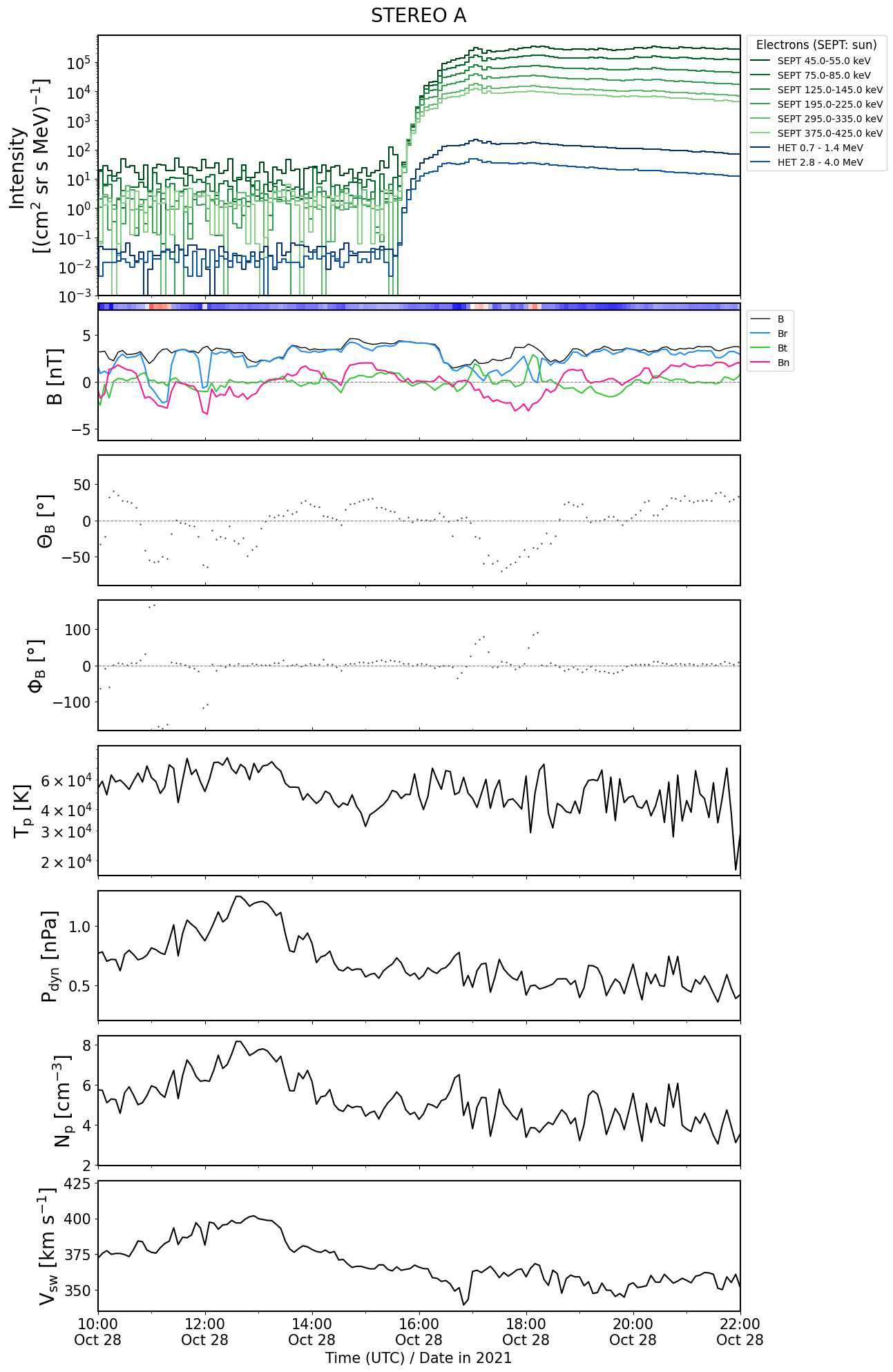} % first figure itself
        \includegraphics[width=0.31\textwidth]{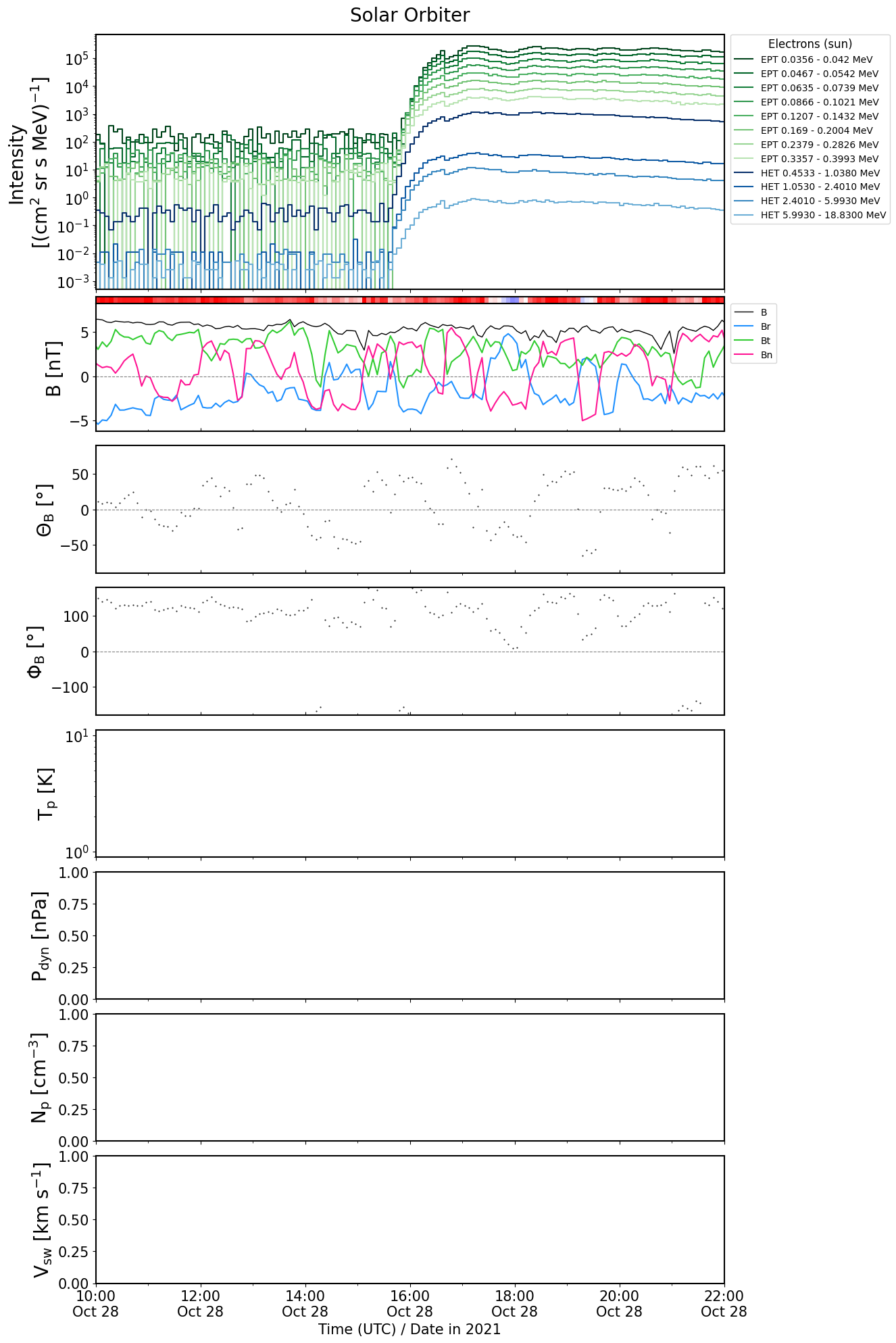} % second figure itself
        \includegraphics[width=0.31\textwidth]{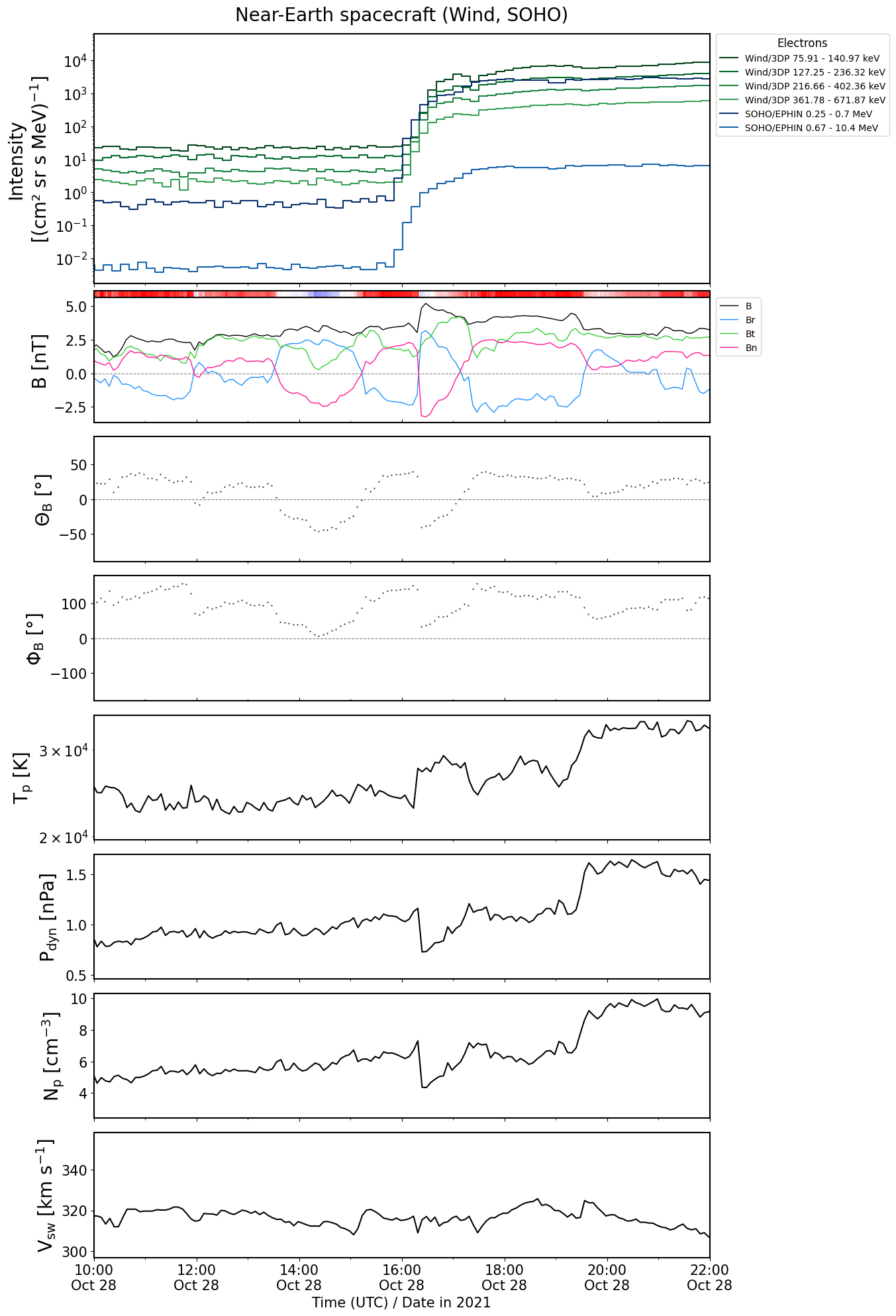} % second figure itself
        \caption{Magnetic field and plasma measurements taken during the SEP event at the observers used in the transport analysis at STA (left), SolOr (middle), and Earth (right) by means of Wind and SOHO data. For each observer, from top to bottom, the panels display the electron fluxes,  magnitude, co-latitudinal and azimuthal angles of the HMF, plasma temperature, dynamic pressure and density, and  velocity of the solar wind. Plasma and solar wind measurements were not available for SolO due to data gap in SWA. }
    \label{fig: field_and_plasma}
\end{figure*}

The plots are generated with the multi-instrument plot tool, provided by SOLER project\footnote{\url{https://github.com/soler-he/sep_tools}} \citep{gieseler_2025_15827561}. In Figure \ref{fig: field_and_plasma}, for each observer, the panels from top to bottom display the electron fluxes, the magnitude, co-latitudinal and azimuthal angles of the HMF, plasma temperature, dynamic pressure and density, and the velocity of the solar wind. Plasma and solar wind measurements were not available for SolO due to data gap in SWA. 

% % %oooooooooooooooooooooooooooooooooooooooooooooooooooooooooooooooooooooooo

 \clearpage
\section{Pitch-angle distributions and anisotropies for electrons detected at STA, SolO, and the Earth}
\label{appenB}

\begin{figure*}[h!]
\vspace{-0.6em}
\centering
\includegraphics[width=0.33\textwidth]{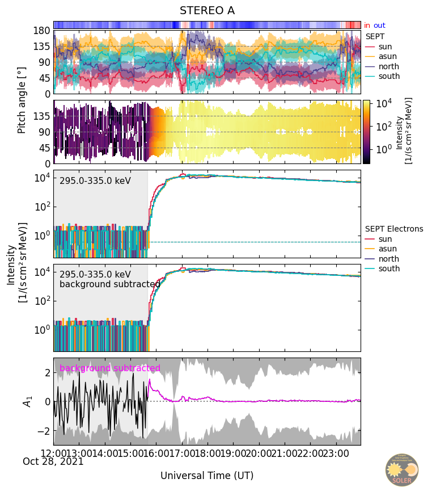}
\includegraphics[width=0.33\textwidth]{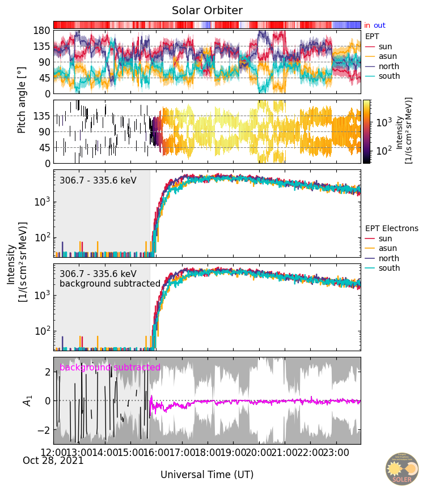}
\includegraphics[width=0.33\textwidth]{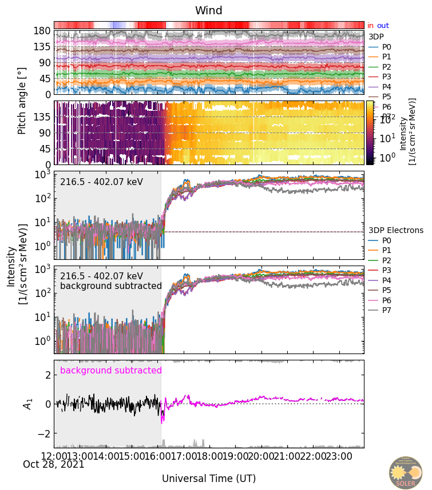}
\vspace{-0.4em}
\caption{PAD and anisotropy for (from left to right) STA/SEPT $\sim$314~keV, SolO/SEPT $\sim$321~keV, and Wind/3DP $\sim$310~keV electrons. In each figure (for each observer), the top panel represents the inward (in red) and outward (in blue) polarity, the second panel depicts the PADs, the third panel shows the sectored intensities in all the available directions, the fourth panel shows the sectored intensities but with background subtraction, and the fifth panel presents the anisotropy. All figures were created with the SEP tools of the SOLER project \citep{gieseler_2025_15827561}.}
\label{fig:pads_314keV}
\vspace{-0.6em}
\end{figure*}

Pitch-angle distributions and anisotropies for $\sim$314~keV electrons detected at STA/SEPT (four sectors), SolO/EPT (four sectors) and Wind/3DP (eight sectors). The corresponding plots in Fig.~\ref{fig:pads_314keV} are created with the SOLER anisotropy tool \citep{gieseler_2025_15827561}. We find that the highest peak intensities and anisotropies are detected at STA, which had the best magnetic connection to the eruption site (CA~$\sim$~25$^{\circ}$). Peak intensity and anisotropy drop with decreasing magnetic connectivity. We further note that STA is located in a mean HMF with a different magnetic polarity than SolO and Earth during the event. This suggests that SEPs had to travel through the HCS to reach all three observers.

\clearpage
\twocolumn
\end{appendix}

%------------------------------------------------------------------------

\end{document}